\documentclass{aa}
\usepackage[varg]{txfonts}
\usepackage{graphicx}
\usepackage{lscape}
\usepackage{txfonts}
\usepackage{color}
\usepackage{natbib}

\begin{document}
%%%%%%%%%%%%%%%%%%%%%%%%%%%%%%%%%%%%%%%%%%%%%%%%

\title{The surface brightness - colour relations based on eclipsing binary stars and calibrated with {\it Gaia} EDR3}
%\thanks{Based on observations collected at the European Southern Observatory in Chile using 3.6m
%telescope and the Swiss 1.2m Euler telescope under
%programmes 084.D-0591, 085.D-0395, 085.C-0614, 086.D-0078, 091.D-0469,
%092.D-0363, and 190.D-0237}
\titlerunning{}
\author{D.~Graczyk\inst{1},
G.~Pietrzy\'nski\inst{2},
C.~Galan\inst{2},
W.~Gieren\inst{3},
A.~Tkachenko\inst{4},
R.I.~Anderson\inst{5,12},
A.~Gallenne\inst{2,3,13},
M.~G{\'o}rski\inst{2},\\
G. Hajdu\inst{2},
M.~Ka{\l}uszy{\'n}ski\inst{2},
P.~Karczmarek\inst{3},
P.~Kervella\inst{8},
P.F.L.~Maxted\inst{9},
N.~Nardetto\inst{7},
W.~Narloch\inst{3},
K.~Pavlovski\inst{10},\\
B.~Pilecki\inst{2},
W.~Pych\inst{2},
J.~Southworth\inst{9},
J.~Storm\inst{11},
K.~Suchomska\inst{2},
M.~Taormina\inst{2},
S.~Villanova\inst{3},
P.~Wielg{\'o}rski\inst{2},\\
B.~Zgirski\inst{2}
\and
P.~Konorski\inst{14}
}
\authorrunning{D. Graczyk et al.}
\institute{Centrum Astronomiczne im. Miko{\l}aja Kopernika, Polish Academy of Sciences, Rabia{\'n}ska 8, 87-100, Toru{\'n}, Poland
\and Centrum Astronomiczne im. Miko{\l}aja Kopernika, Polish Academy of Sciences, Bartycka 18, 00-716, Warsaw, Poland
\and Departamento de Astronom{\'i}a, Universidad de Concepci{\'o}n, Casilla 160-C, Concepci{\'o}n, Chile
\and Instituut voor Sterrenkunde, KU Leuven, Celestijnenlaan 200D, B-3001 Leuven, Belgium
\and European Southern Observatory, Karl-Schwarzschild-Str. 2, 85748 Garching b. M{\"u}nchen, Germany
\and European Southern Observatory, Alonso de C{\'o}rdova 3107, Casilla 19001, Santiago, Chile
\and Universit\'e C$\hat{\rm o}$te d'Azur, Observatoire de la C$\hat{\rm o}$te d'Azur, CNRS, Laboratoire Lagrange, Nice, France
\and LESIA, Observatoire de Paris, Universit\'e PSL, CNRS, Sorbonne Universit\'e, Universit\'e de Paris, 5 place Jules Janssen, 92195 Meudon, France
%\and Laboratoire Lagrange, UMR7293, Universit\acute{e} de Nice Sophia-Antipolis, CNRS, Observatoire de la C\hat{o}te d?Azur, Nice, France
\and Astrophysics Group, Keele University, Staffordshire, ST5 5BG, UK
\and Department of Physics, University of Zagreb, Bijeni{\'c}ka cesta 32, 10000 Zagreb, Croatia
%\and Warsaw University Observatory, Al. Ujazdowskie 4, 00-478 Warsaw, Poland
\and Leibniz-Institut f\"{u}r Astrophysik Potsdam, An der Sternwarte 16, 14482 Potsdam, Germany
\and Institute of Physics, Laboratory of Astrophysics, Ecole Polytechnique F{\'e}d{\'e}rale de Lausanne (EPFL), Observatoire de Sauverny, 1290 Versoix, Switzerland
\and Unidad Mixta International Franco-Chilena de Astronom{\'i}a (CNRS UMI 3386), Departamento de Astronom{\'i}a, Universidad de Chile, Camino El Observatorio 1515, Las Condes, Santiago, Chile
\and LunAres Research Station, Airport, Hangar 11, Pi{\l}a, Poland
%\and Observatoire de Gen{\`e}ve, Universit{\'e} de Gen{\`e}ve, 51 Ch. des Maillettes, 1290 Sauverny, Switzerland
%\and Department of Physics and Astronomy, Johns Hopkins University, Baltiore, MD 21218, USA
%\and Department of Physics, University of Warwick, Coventry CV4 7AL, UK
}
%\mail{darek@astro-udec.cl}
%ESO ID: 085.C-0614, 085.D-0395, 086.D-0078, 087.C-0012, 089.C-0415, 190.D-0237, 091.D-0469

%\date{Accepted ... . Received ...; in original form ...}

%\pagerange{\pageref{firstpage}--\pageref{lastpage}} \pubyear{2014}

%\label{firstpage}

\abstract{} 
{The surface brightness -- colour relation (SBCR) is a basic tool in establishing precise and accurate distances within the Local Group. Detached eclipsing binary stars with accurately determined radii and trigonometric parallaxes allow for a calibration of the SBCRs with unprecedented accuracy.} 
{We analysed four nearby eclipsing binary stars containing late F-type main sequence components: AL~Ari, AL~Dor, FM~Leo and BN~Scl. We determined very precise spectroscopic orbits and combined them with high precision ground- and space-based photometry. We derived the astrophysical parameters of their components with mean errors of 0.1\% for mass and 0.4\% for radius. We combined those four systems with another 24 nearby eclipsing binaries with accurately known radii from the literature for which \textit{Gaia} EDR3 parallaxes are available, in order to derive the SBCRs.} 
{The resulting SBCRs cover stellar spectral types from B9\,V to G7\,V. For calibrations we used Johnson optical $B$ and $V$, \textit{Gaia} $G_{\rm BP}$ and $G$ and 2MASS $JHK$ bands. The most precise relations are calibrated using the infrared $K$ band and allow to predict angular diameters of  A-, F-, and G-type dwarf and subgiant stars with a precision of 1\%.}
{}

%Stellar rotation of both components is already pseudo-synchronized during periastron passage.


\maketitle

\section{Introduction}

The concept of the stellar surface brightness parameter $S\!$ is useful in astrophysics because it connects the stellar absolute magnitude $M$ with the stellar radius $R$ by a very simple relation \citep{wes69}. It is very convenient to express the $S\!$ parameter as a function of an intrinsic stellar colour, called the surface brightness -- colour relation (SBCR), giving a powerful tool in predicting the angular diameters of stars \citep[e.g.][]{bar76,VBe99,ker04}. When the distance (or the trigonometric parallax) to a particular star is known an application of an SBCR immediately gives its radius \citep{lac77a}. When, on the other hand, the radius of a star is known, an application of SBCR gives a robust distance \citep{lac77b}. The latter approach, in particular, has resulted in very precise distance determinations to the Magellanic Clouds \citep[e.g.][]{pie19,gra20}, setting the zero-point of the extragalactic distance ladder with a precision of $\sim\!1$\%.

% In this paper we present the trigonometric parallax to the stellar surface brightness inference based on eclipsing binary stars.
In this paper we present a calibration of SBCRs based on eclipsing binary stars with trigonometric parallaxes.
The fundamentals of the method are given in \cite{gra17}. The first application of the method was presented by \cite{ste10} in the case of Algol, but for many years a lack of sufficient-quality parallaxes of nearby eclipsing binaries compromised a successful calibration of SBCRs this way. The results of the \textit{Hipparcos} space mission \citep{per97} showed that there are many nearby ($d<200$ pc) eclipsing binaries suitable for the SBCR calibration. The best systems are those with well detached, spotless components of similar temperature and size, and with a low interstellar reddening. \cite{kru99} published a list of potential targets and during subsequent years we have obtained photometric and spectroscopic follow-up of about 50 eclipsing binaries based mostly on their list.  The analysis of the \textit{Hipparcos} sample proved to be difficult: many of the systems turned out to be triples or active systems not suitable for the SBCR calibration, photometric follow-up with a sufficiently high quality was tedious and many parallaxes were of a low precision. As a result we have so far published a full analysis of only two eclipsing binaries: IO~Aqr \citep{gra15} and LL~Aqr \citep{gra16}. Also for a few systems we derived the orbital parallaxes and precise masses but we assumed radii from literature \citep{gal16,gal19}. However, with the advent of the space missions {\it Gaia} \citep{gaia16}, {\it Kepler} \citep{koch10} and TESS \citep{ric15} the situation has improved dramatically. Here we present an analysis of four eclipsing binary stars and use {\it Gaia} parallaxes to obtain a very precise SBCR calibration.

\section{Observations}\label{observ}

\subsection{Sample of stars}

AL~Ari, AL~Dor, FM~Leo and BN~Scl were discovered as variable stars during the \textit{Hipparcos} space mission \citep{per97}, classified as eclipsing binaries and given names in the General Catalogue of Variable Stars (GCVS) by \cite{kaz99}. Because the systems are well-detached, close to the Sun and have no significant spot activity, we included them in our \textit{Hipparcos} sample.

AL~Ari was selected as a follow-up target in a photometric and spectroscopic survey of solar-type eclipsing binaries by \cite{cla01}. The preliminary physical parameters of this system were included in a compilation of eclipsing binaries prepared by \cite{gra17}. AL~Dor was included in a sample of eclipsing binaries used by \cite{gra19} to investigate the zero-point of the \textit{Gaia} parallaxes. Precise masses of its components and an interferometric orbital parallax were derived by \cite{gal19}. Here we present the full analysis of those two systems with updated physical parameters. An analysis of light- and radial velocity curves of FM~Leo was published by \cite{rat10}, who derived the fundamental physical parameters of the system. Updated mass measurements of the components, together with the analysis of the Rossiter-McLaughlin effect, were provided by \cite{syb18}. The last of the systems, BN~Scl, has not so far been studied in detail. General data on the four systems are given in Table~\ref{tab:basic}.

\subsection{Photometry}\label{photo}

\subsubsection{Ground-based Str{\"o}mgren photometry}

We used Str{\"o}mgren $uvby$ photometry of AL~Ari secured with the Str\"omgren Automated Telescope (SAT) at ESO, La Silla \citep{cla01}. The photometry was kindly provided to us by E.~Olsen. The data were taken between October 1997 and December 1998 and comprise 827 differential magnitudes with respect to three comparison stars (HD 15814, 16707 and 17499) in each filter. The photometry was detrended and normalised separately in each filter -- see Table~\ref{tab:alari}.

\subsubsection{Space-based photometry}

AL~Dor was observed by the TESS space mission \citep{ric15} in short cadence during sectors 1--13 and 27--29. We downloaded a total of 248,078 photometric data points collected by TESS from the Mikulski Archive for Space Telescopes (MAST) archive. The sectors differ somewhat in the number of artefacts and outliers, instrumental drifts, and observational noise. We decided to use only sectors 1--3, 12 and 27 in our analysis, as they have the smallest drifts and out-of-eclipse {\it rms} and cover a long time interval. Depending on the number of outliers and artefacts we used the Simple Aperture Photometry (\verb"SAP_FLUX") or the Pre-search Data Conditioning SAP (\verb"PDCSAP_FLUX") fluxes. The eclipse depths differ slightly between sectors, by up to 0.5\% in flux. Only in the first five sectors is the depth constant to within 0.1\%, and we renormalised the light curves from sectors 12 and 27 to reach the same minima depths. We retained all data during and close to eclipses, but kept only every 50th datapoint from the out-of-eclipse parts of the light curve. In total we used 6387 data points in our analysis below.

FM~Leo was observed by the K2 mission \citep{how14}, the extension of the {\it Kepler} space mission \citep{koch10}. It was observed during campaign 1 in 2014, in both long and short cadence.  We downloaded these data from the MAST archive, containing 3192 long-cadence and 53,048 short-cadence datapoints. The data were detrended with a third order order spline function to remove only long-term flux changes and to keep the small ellipsoidal out-of-eclipse variations. The data were cleaned of obvious outliers and the fluxes were subsequently normalised. In order to lower the number of datapoints we removed short-cadence data taken outside eclipse and replaced them with the long-cadence data. This resulted in 6433 photometric points: 3756 short-cadence points around eclipses and 2677 long-cadence points outside eclipses.

BN~Scl was observed by TESS in sectors 1 and 29. In both sectors some spot activity on the components can be noted, especially in the first part of the light curve from sector 1. The out-of-eclipse flux changes due to the spots are smaller than 0.5\%, but they do slightly affect the shape and depth of the minima. From sector 29 we chose the data obtained in the time interval between BJD 2459088 and 2459100, when the light curve is most symmetric around both eclipses, i.e.\ the fluxes were minimally influenced by the spots. We used the \verb"PDCSAP_FLUX" fluxes and retained points around eclipses and every second point in the out-of-eclipse parts of the light curve. In total we ended up with 3521 datapoints.

\subsection{Spectroscopy}

\subsubsection{HARPS}
\label{harps}

We obtained spectra of the four systems with the High Accuracy Radial velocity Planet Searcher \citep[HARPS;][]{may03} on the European Southern Observatory 3.6-m telescope in La Silla, Chile. In total we collected 64 spectra between 2009 February 26 and 2015 June 21. The targets are bright and typical integration times were shorter than 10 minutes. The average signal-to-noise ratio (S/N) of the spectra was 75 for AL~Dor, FM~Leo and BN~Scl, and 40 for AL~Ari. All spectra were reduced on-site using the HARPS Data Reduction Software (DRS).

\subsubsection{CORALIE}

30 spectra were obtained for BN~Scl and AL~Dor with the CORALIE spectrograph on the Swiss 1.2-m Euler Telescope at La Silla between 2008 October 2 and 2010 November 1. The exposure times were between 800\,s and 1100\,s for BN Scl while for AL Dor they were between 520\,s and 1100\,s depending on the sky conditions and the airmass. A typical S/N near 5500\AA\ was 20 per pixel. The spectra taken during 2009 and 2010 were secured with a simultaneous ThAr lamp exposure, allowing for correcting for the intra-night radial velocity drifts, while spectra taken during 2008 not. That may add some scatter to the radial velocities derived from CORALIE 2008 spectra.The spectra were reduced using the Geneva pipeline, which does all the standard steps.

\begin{table*}
\begin{center}
\caption{Basic data on the eclipsing binary stars.}
\label{tab:basic}
\begin{tabular}{lcccc}
\hline \hline
Parameter & AL~Ari & AL~Dor & FM~Leo & BN~Scl \\
\hline
R.A. (2000) & $02^{\rm h} 42^{\rm m} 36\fs34$&$04^{\rm h} 46^{\rm m} 52\fs26$ &$11^{\rm h} 12^{\rm m} 45\fs09$ &$23^{\rm h} 07^{\rm m} 42\fs75$\\
Dec (2000)& $+12\degr 44\arcmin 07\farcs8$&$-60\degr 36\arcmin 12\farcs7$  &$+00\degr  20\arcmin 52\farcs8$  &$-30\degr 14\arcmin 00\farcs0$ \\
$\varpi_{Gaia/EDR3}$ (mas) & 7.236$\pm$0.021& 15.162$\pm$0.016& 6.831$\pm$0.026 &5.480$\pm$0.016 \\
Sp. Type &F6V+G7V&F9V+F9V&F6V+F6V&F7V+F8V\\
$V$ (mag) &  9.236$\pm$0.020& 7.738$\pm$0.017& 8.471$\pm$0.018& 8.931$\pm$0.018\\
 $B\!-\!V$ (mag)& 0.520$\pm$0.027& 0.566$\pm$0.023&0.492$\pm$0.027& 0.489$\pm$0.015\\
$P_{\rm obs}$ (days) &3.74745&14.9054&6.72861&3.65053 \\
\hline
\end{tabular}
\end{center}
\end{table*}

\begin{table*}
\centering
\caption{The $uvby$ photometry of AL Ari.  The full data will be available at CDS.}
\label{tab:alari}
\begin{tabular}{@{}ccccc@{}}
\hline \hline
Date & \multicolumn{4}{c}{Normalized Flux} \\
 HJD $-$ 2450000 & $u$ & $v$ & $b$ & $y$ \\
\hline
741.71175 &0.99675 &0.99522 &0.99934 &0.99939 \\
741.71762 &0.99400 &0.99706 &0.99566 &0.99480 \\
741.72142 &0.98127 &0.97976 &0.98110 &0.97934 \\
742.70690 &0.97406 &0.97976 &0.97121 &0.97305 \\
742.71239 &0.99767 &1.00443 &1.00395 &0.99755 \\
742.71635 &0.99126 &0.98610 &0.98563 &0.99023 \\
\hline
\end{tabular}
\end{table*}

\section{Analysis of spectra}
\label{analys}
\subsection{Radial velocities \label{rad_vel}}

\begin{table*}
\centering
\caption{RV measurements for eclipsing binary stars. The full data will be available at CDS.}
\label{tab_rv}
\begin{tabular}{@{}llccccccc@{}}
\hline \hline
Object & BJD & $RV_1$ & $RV_1$ error & $RV_2$ & $RV_2$ error & Spectrograph \\
 & -2450000& (km s$^{-1}$) & (km s$^{-1}$) & (km s$^{-1}$) &(km s$^{-1}$) & \\
\hline
AL~Ari & 5447.69694 &   58.562   &0.061 &$-$118.396   &0.154&  HARPS \\
AL~Ari & 5447.77210  &  58.802   &0.057 &$-$118.906   &0.136 & HARPS \\
AL~Ari & 5449.70066   &$-$91.409 &  0.055&   73.145 &  0.134&  HARPS \\
AL~Ari & 5467.70349   &$-$63.990  & 0.058 &  38.197  & 0.145 & HARPS \\
AL~Ari & 5468.92022   &$-$56.462   &0.070  & 28.700  &0.141 & HARPS \\
\hline
\end{tabular}
\end{table*}

\begin{figure}
\hspace*{-0.5cm}
\includegraphics[angle=0,scale=0.57]{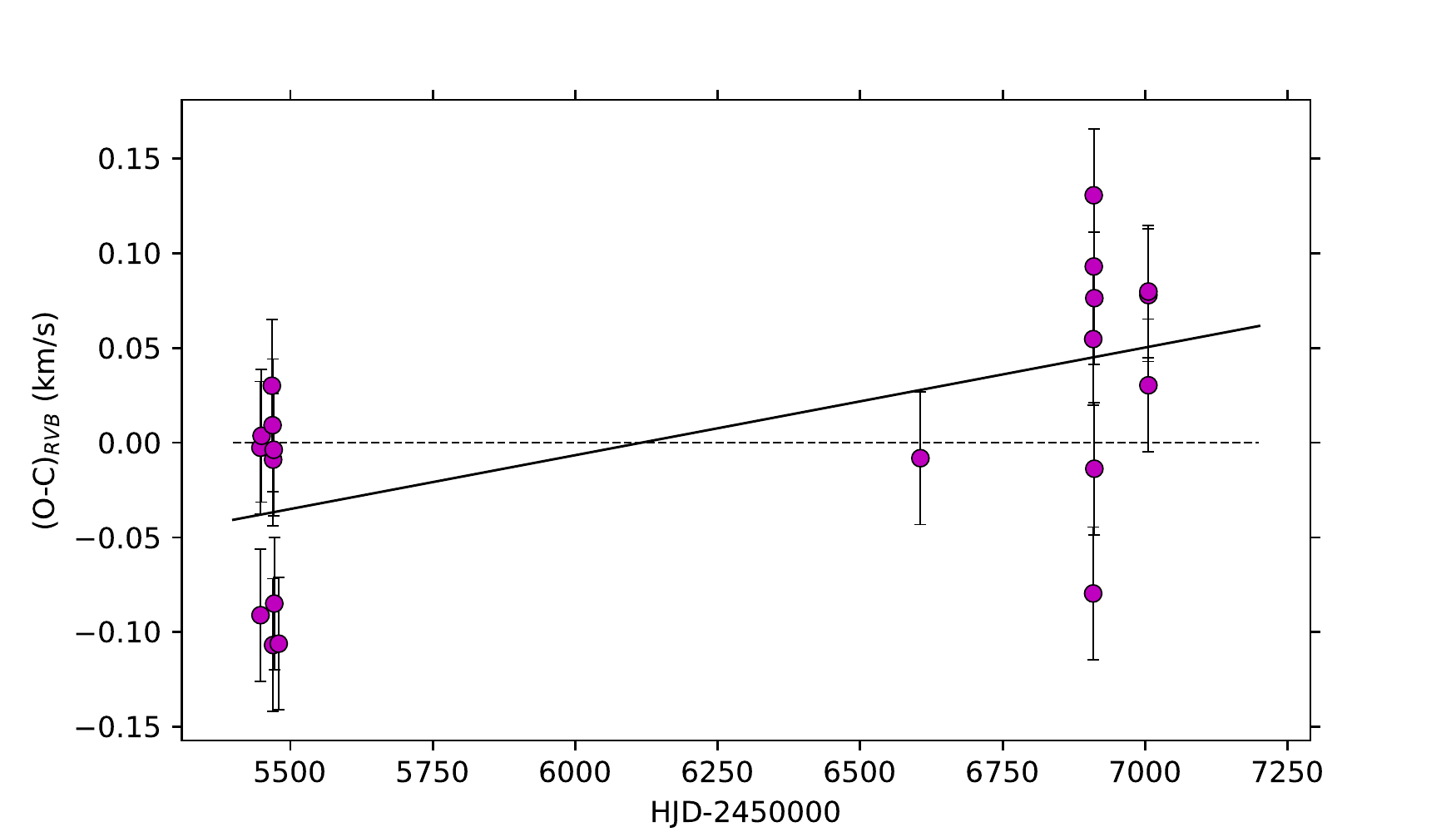}
\caption{The barycentric radial velocity trend in AL~Ari. The continuous line is a simple linear function fit to the trend.\label{rv:trend}}
\end{figure}

\label{sec:rv}
We used the RaveSpan code \citep{pil17} to measure the radial velocities of the components in all four systems via the Broadening Function (BF) formalism \citep{ruc92,ruc99}. We used templates from the library of synthetic LTE spectra by \cite{col05} matching the mean values of estimated effective temperature and gravity of the stars in the binary. The abundances were assumed to be solar. The line profiles of the components of AL~Ari, AL~Dor and FM~Leo are Gaussian and suggest small rotational velocities, while the line profiles of the components of BN~Scl are clearly rotationally broadened. The typical precision of radial velocity determination was about 20--30~m~s$^{-1}$ for AL~Dor and FM~Leo and about 50--100~m~s$^{-1}$ for the faster-rotating AL~Ari and BN~Scl. With RaveSpan we made initial radial velocity solutions to the data and noted a small systematic drift in the $O\!-\!C$ radial velocity residuals of AL~Ari (see Fig.~\ref{rv:trend}). We fitted a linear function to the observed residuals and derived a radial velocity acceleration of the AL~Ari barycentre of $0.21\pm0.04$ m~s$^{-1}$ per orbital period (cycle). The trend was removed before we made full simultaneous light- and radial velocity curve analysis of AL~Ari. The radial velocity measurement are summarised in Table~\ref{tab_rv}.

\subsection{Spectra disentangling\label{deco}}

We disentangled the observed composite spectra into those of the individual components of the binaries using an iterative method outlined by \cite{gon06} and implemented in the RaveSpan code. This was done only for the HARPS spectra due to their higher S/N. We used the radial velocities previously measured and iterated twice. In order to account for each star's light contribution to the system we renormalised the spectra using the photometric parameters from our simultaneous light- and radial velocity curve solutions (see Sec.~\ref{wd}).  The resulting individual spectra have S/N values at 5500 $\AA$ of about 100, except for that of the secondary component of AL~Ari which has S/N $=$ 40.

\subsection{Stellar atmospheric analysis}\label{abu}

\subsubsection{Methods}

We used the high-resolution disentangled HARPS spectra to derive the atmospheric parameters of the components of the binary systems. For this we employed the `Grid Search in Stellar Parameters' ({\sl GSSP}) software package, designed for the analysis of high-resolution spectra of single stars and binary systems \citep{Tka2015}. We used the latest version of {\sl GSSP} which makes use of the most recent (August 2020) version of the radiative transfer code, including a better solver for molecular transitions and improved calculation of the continuum, as well as a new, updated grid of atmosphere models. The code uses the spectrum synthesis method by employing the {\sl S\textsc{ynth}V} LTE-based radiative transfer code \citep{Tsy1996}.  We used a grid of atmosphere models that were calculated with the use of the {\sl LL\textsc{models}} code \citep{Shu2004}. Important advantages of {\sl GSSP} include its ability to perform fast calculations for large grids of synthetic models through the use of parallel computing via the {\sl O\textsc{pen}MPI$\,$} implementation\footnote{http://www.open-mpi.org} and an accurate treatment of limb darkening by computing the specific line and continuum intensities for different positions on the stellar disk.
%{\it The latter is important in addressing the issue of the calibration of the surface brightness of stars in binary systems.}

Our approach relies on the use of the {\sl single} and {\sl binary} versions of the code.  We aim to determine the surface atmospheric parameters such as the effective temperature ($T_{\rm{eff}}$), the surface gravity ($\log{g}$), the metallicity ([M/H]), and the microturbulent velocity ($\xi$). More detailed investigation of surface abundances of individual chemical elements are scheduled for a follow-up paper. The {\sl GSSP}$\_${\sl single} module is designed for the analysis of the spectra of single stars or the disentangled spectra of the individual components of double-lined spectroscopic binary systems.  To use this option the disentangled spectra should be renormalised appropriately taking into account the wavelength dependence of the light dilution factor ($f_{\rm i}$). Otherwise, as the single version uses only the wavelength-independent $f_{\rm i}$, the results will be distorted the more the components of the system differ in $T_{\rm eff}$.

The {\sl GSSP}$\_${\sl binary} module is designed for dealing simultaneously with disentangled spectra of both components by taking into account the wavelength dependence of $f_{\rm i}$, which is defined as the fractional contribution of the individual components to the total light of the entire system \citep[see][for details]{Tka2015}.  For our purposes, we used the 'fit' option to search for the best matching among the grid of synthetic spectra to the observed ones.  Both {\sl GSSP} modules calculate synthetic spectra for each set of grid parameters and compares them with the normalised observed spectrum. The $\chi^2$ function is calculated to judge the goodness of fit. The code searches for the minimum $\chi^2$, and provides the best-fit parameters together with the values of reduced $\chi^2$ 
%($\chi^2$ value normalised to the number of pixels in the spectrum, minus the number of free parameters)
as well as $1\sigma$ uncertainty level in terms of $\chi^2$.

In our disentangled HARPS spectra the continuum levels are not reconstructed well enough in the region of strong hydrogen lines, characterised by the presence of wide wings.  Therefore the appropriate regions of H$\alpha$, H$\beta$, and H$\gamma$ lines are excluded from the analysis.  The `blue' region at wavelengths shorter than the H$\gamma$ line is excluded as well due to a significantly lower S/N.  We also skipped the region between $\sim 6274$ -- $6282$\,\AA\, contaminated by oxygen molecules in Earth's atmosphere.  The regions used for fitting of the synthetic models are shown in Table\,\ref{T_WRs}.
%mb-finish

\begin{table}
\centering
\caption{Wavelength ranges used in $\chi^2$ calculations.}
\label{T_WRs}
\begin{tabular}{@{}l@{\hskip 20mm}l@{}}
\hline \hline
Starting value (\AA) & Ending value (\AA) \\
\hline
4385 & 4810 \\
4910 & 5298 \\
5344 & 6274 \\
6282 & 6500 \\
6630 & 6780 \\
\hline
\end{tabular}
\end{table}

\begin{table*}[h!]
\centering
\caption{The best-fitting final parameters of model atmospheres matching the observed spectra.
The simultaneous solution obtained with
the {\sl GSSP}$\_${\sl binary} module for each eclipsing binary using the disentangled spectra of both components.
%The best fitcparameters together with the uncertainties estimated using thecreduced $\chi^2$ and the 1\,$\sigma$ level in $\chi^2$ ($\chi^2_{1\sigma}$).
}
\label{T_atm_par}
\begin{tabular}{@{}l@{\hskip 24mm}l@{\hskip 12mm}l@{\hskip 24mm}l@{\hskip 12mm}l@{\hskip 24mm}l@{}}
\hline \hline
\multicolumn{6}{c}{AL~Ari}\\
\hline
                       & \multicolumn{2}{l}{~Grid best model}& \multicolumn{2}{l}{~~~~~~Best fit parameters} &          \\
Parameter              & Primary  & Secondary                      & Primary             & Secondary           & Unit         \\
\hline
$[$M/H$]$              & $-0.4$   & $-0.6$                         & $-0.42\pm0.07$      & $-0.60\pm0.13$      & dex          \\
$T_{\rm{eff}}$         & $6600$   & $5600$                         & $6578\pm82$         & $5560\pm192$        & K            \\
$\log{g}^{\ast}$       & $4.2$    & $4.5$                          & $4.30\pm0.15$       & $4.26\pm0.37$       & dex          \\
$\xi$                  & $1.5$    & $2.0$                          & $1.54\pm0.14$       & $2.00\pm0.46$       & km\,s$^{-1}$ \\
$\zeta^{\star}$        & $5.8$    & $2.6$                          & $5.8^{+1.6}_{-2.0}$ & $2.6^{+3.0}_{-1.1}$ & km\,s$^{-1}$ \\
$V_{\rm{rot}} \sin{i}$ & $17$     & $13$                           & $17.11\pm0.61$      & $12.65\pm1.50$      & km\,s$^{-1}$ \\
reduced $\chi^2$       & \multicolumn{3}{c}{8.304440}              &                                           & -            \\
$\chi^2_{1\sigma}$     & \multicolumn{3}{c}{8.349265}              &                                           & -            \\
\multicolumn{6}{c}{}\\
\hline \hline
\multicolumn{6}{c}{AL~Dor}\\
\hline
                       & \multicolumn{2}{l}{~Grid best model}& \multicolumn{2}{l}{~~~~~~Best fit parameters} &          \\
Parameter              & Primary  & Secondary                      & Primary             & Secondary           & Unit         \\
\hline
$[$M/H$]$              & $-0.1$   & $-0.1$                         & $-0.11\pm0.05$      & $-0.11\pm0.05$      & dex          \\
$T_{\rm{eff}}$         & $6100$   & $6100$                         & $6076\pm70$         & $6051\pm70$         & K            \\
$\log{g}^{\ast}$       & $4.4$    & $4.4$                          & $4.38\pm0.18$       & $4.38\pm0.17$       & dex          \\
$\xi$                  & $1.4$    & $1.3$                          & $1.36\pm0.18$       & $1.29\pm0.17$       & km\,s$^{-1}$ \\
$\zeta^{\star}$        & $3.1$    & $3.4$                          & $3.1\pm0.5$         & $3.4\pm0.5$         & km\,s$^{-1}$ \\
$V_{\rm{rot}} \sin{i}$ & $4$      & $4$                            & $4.08\pm0.60$       & $3.90\pm0.66$       & km\,s$^{-1}$ \\
reduced $\chi^2$       & \multicolumn{3}{c}{16.507021}             &                                           & -            \\
$\chi^2_{1\sigma}$     & \multicolumn{3}{c}{16.596123}             &                                           & -            \\
\multicolumn{6}{c}{}\\
\hline \hline
\multicolumn{6}{c}{FM\,Leo}\\
\hline
                       & \multicolumn{2}{l}{~Grid best model}& \multicolumn{2}{l}{~~~~~~Best fit parameters} &          \\
Parameter              & Primary  & Secondary                      & Primary             & Secondary           & Unit         \\
\hline
$[$M/H$]$              & $-0.1$   & $-0.1$                         & $-0.10\pm0.06$      & $-0.10\pm0.06$      & dex          \\
$T_{\rm{eff}}$         & $6400$   & $6400$                         & $6425\pm88$         & $6418\pm96$         & K            \\
$\log{g}^{\ast}$       & $4.1$    & $4.2$                          & $4.11\pm0.17$       & $4.18\pm0.19$       & dex          \\
$\xi$                  & $1.6$    & $1.5$                          & $1.61\pm0.15$       & $1.49\pm0.17$       & km\,s$^{-1}$ \\
$\zeta^{\star}$        & $5.8$    & $4.3$                          & $5.8^{+1.0}_{-1.3}$ & $4.3^{+1.3}_{-1.6}$ & km\,s$^{-1}$ \\
$V_{\rm{rot}} \sin{i}$ & $11$     & $11$                           & $11.0\pm0.6$        & $11.0\pm0.6$        & km\,s$^{-1}$ \\
reduced $\chi^2$       & \multicolumn{3}{c}{17.056393}             &                                           & -            \\
$\chi^2_{1\sigma}$     & \multicolumn{3}{c}{17.148460}             &                                           & -            \\
\multicolumn{6}{c}{}\\
\hline \hline
\multicolumn{6}{c}{BN\,Scl}\\
\hline
                       & \multicolumn{2}{l}{~Grid best model}& \multicolumn{2}{l}{~~~~~~Best fit parameters} &          \\
Parameter              & Primary  & Secondary                      & Primary             & Secondary           & Unit         \\
\hline
$[$M/H$]$              & $-0.2$   & $-0.3$                         & $-0.18\pm0.08$      & $-0.24\pm0.10$      & dex          \\
$T_{\rm{eff}}$         & $6300$   & $6300$                         & $6304\pm93$         & $6296\pm148$         & K            \\
$\log{g}^{\ast}$       & $4.0$    & $4.3$                          & $4.15\pm0.18$       & $4.37\pm0.30$       & dex          \\
$\xi$                  & $1.6$    & $1.7$                          & $1.62\pm0.17$       & $1.69\pm0.30$       & km\,s$^{-1}$ \\
$\zeta^{\star}$        & $5$      & $5$                            & $2.8^{+3.9}_{-2.8}$ & $1.0^{+4.5}_{-1.0}$ & km\,s$^{-1}$ \\
$V_{\rm{rot}} \sin{i}$ & $24$     & $19$                           & $23.55\pm1.04$      & $18.73\pm1.43$      & km\,s$^{-1}$ \\
reduced $\chi^2$       & \multicolumn{3}{c}{13.856710}             &                                           & -            \\
$\chi^2_{1\sigma}$     & \multicolumn{3}{c}{13.931628}             &                                           & -            \\
\multicolumn{6}{c}{}\\
\hline
\multicolumn{6}{p{15cm}}{\small{$\ast$ -- fixed on the values available in the grid of atmosphere models that are the closest to the values originated from the dynamical solution.}}\\
\multicolumn{6}{p{15cm}}{\small{$\star$ -- fixed on the values resulting from solutions with the use of {\sl GSSP}$\_${\sl single} module.}}\\
\hline
\end{tabular}
\end{table*}

\subsubsection{Atmospheric parameters}
\label{temp:atm}

As input parameters we used $T_{\rm eff}$ values derived from colours (Sect.~\ref{temp:col}) and the ratio of the stellar radii from the dynamical solution with the Wilson-Devinney code \citep[][hereafter WD]{wil71}. To take into account the turbulent motions in the stellar atmosphere on various scales, we need information about two additional parameters -- the microturbulent ($\xi$) and macroturbulent ($\zeta$) velocities. Their values were estimated using known correlations between turbulence, $\log{g}$, and spectral type or $T_{\rm{eff}}$ \citep{Gray2001, Gray2005, Sma2014, She2019}.  For the metallicity $[$M/H$]$ the solar values were initially adopted.

We started this analysis with the {\sl single} module of the {\sl GSSP} code, utilising disentangled spectra corrected (renormalised) for the dilution factor. We were interested to find in this way the value of the macroturbulent velocity ($\zeta$), which is not allowed to be treated as a free parameter in the {\sl binary} mode.
% because of the strong correlation with the other velocity parameters it is degenerated in most ($\sim99\%$) cases.
The surface gravities were fixed to the values from the dynamical solution (see Section~\ref{WD_results}). The five free parameters were $T_{\rm{eff}}$, $[$M/H$]$, $\xi$, $\zeta$ and $V_{\rm{rot}} \sin{i}$.  In the beginning we used coarse but wide grids of fitted parameters to find the region around the global minimum. In the next steps, the parameter ranges were gradually narrowed and the sampling was finer.  This way we found the solution in several iterations for the best-matching models corresponding to the requested atmospheric parameters.

The above procedure was repeated with the use of the {\sl GSSP}$\_${\sl binary} module applied now to
uncorrected spectra of both components simultaneously.  The values of $\zeta$ were fixed on
values found with the {\sl single} module. The ratio of the components' radii ($R_{1}/R_{2}$), needed
for the code to calculate the wavelength-dependent dilution factor, was
obtained from the fractional radii of the components (see Section~\ref{WD_results}).

\begin{figure*}
\begin{minipage}[th]{0.5\linewidth}
\includegraphics[width=\textwidth]{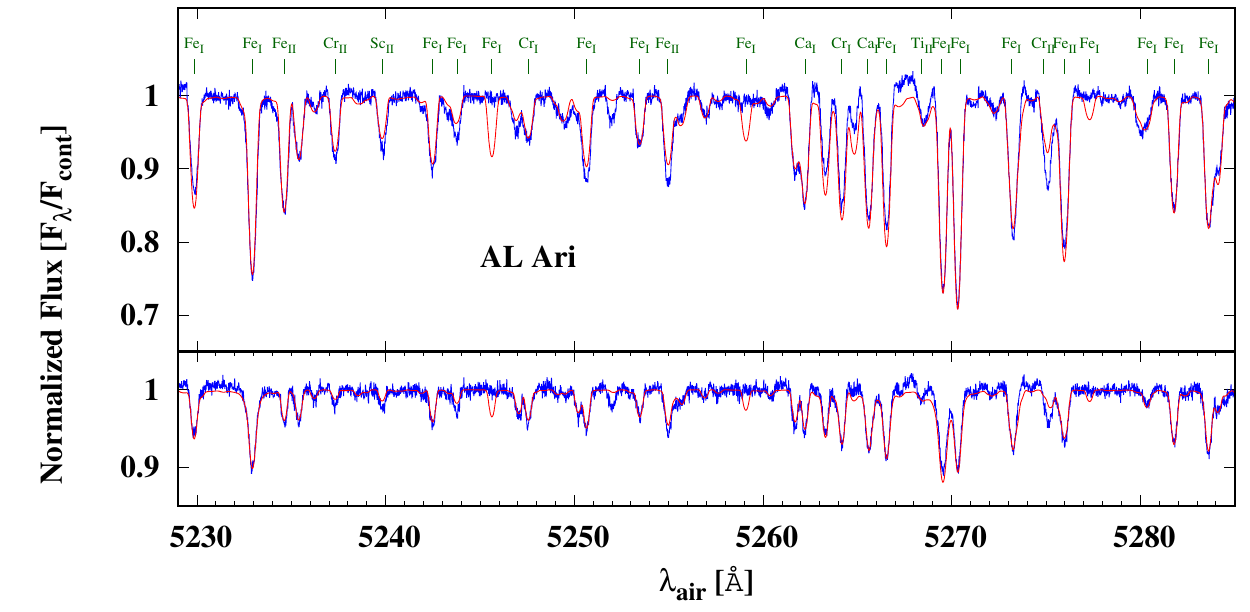} \vspace{-0.15cm}
\mbox{}
\includegraphics[width=\textwidth]{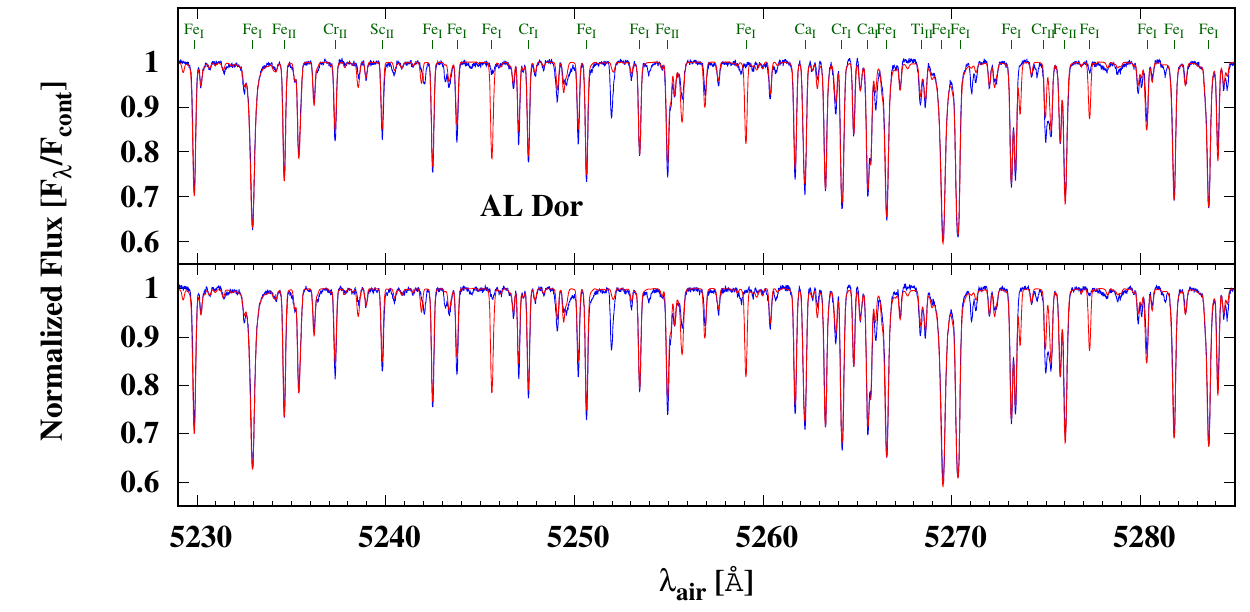}
\end{minipage}\hfill
\begin{minipage}[th]{0.5\linewidth}
\includegraphics[width=\textwidth]{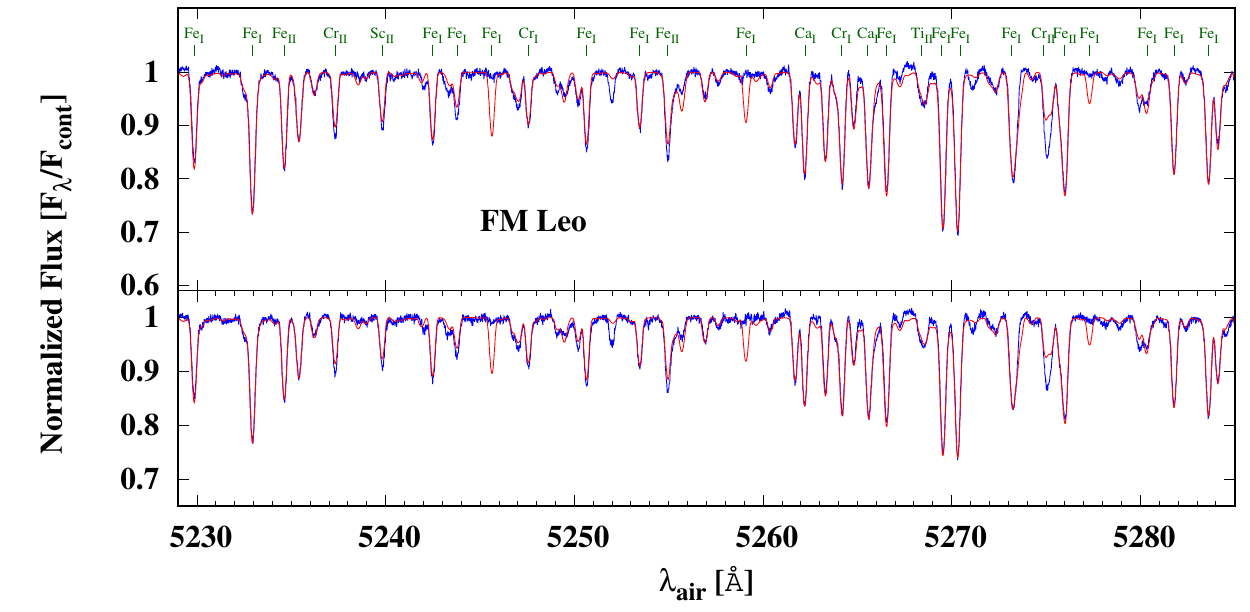} \vspace{-0.15cm}
\mbox{}
\includegraphics[width=\textwidth]{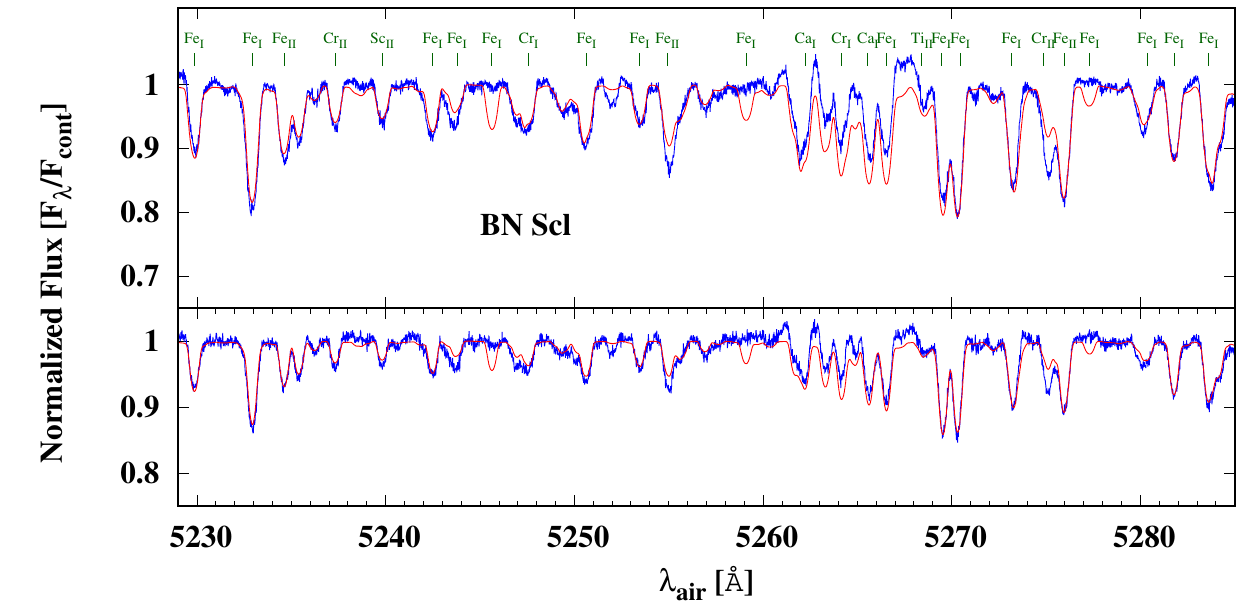}
\end{minipage}\hfill
\caption{The 5229--5285 \AA\, region of uncorrected disentangled spectra of the primary (top)
and the secondary (bottom) components of AL~Ari, AL~Dor, FM\,Leo and BN\,Scl with selected spectral
lines identified. The red lines denote the atmosphere models fits to the observed spectra (blue). The rotational broadening of the absorption lines is easily visible in AL~Ari and BN~Scl.}
\label{F_5229-5286}
\end{figure*}

The 1$\sigma$ uncertainties have been estimated by finding the intersection of the 1$\sigma$ levels in $\chi^2$ with the polynomial functions that have been fitted to the values of reduced $\chi^2$ as recommended by \citet{Tka2015}. In general the {\sl binary} module produced more consistent results, however it depends on how much the atmospheric parameters of the components differ. For example, in the case of AL~Dor where the components are almost indistinguishable, we obtained a perfect match of the results produced with the {\sl single} and {\sl binary} versions. The resulting final parameters from {\sl binary} module are shown in Table\,\ref{T_atm_par}. The samples of disentangled spectra in the region 5229 -- 5285 \AA\, are compared with the best fit synthetic spectra in Figure\,\ref{F_5229-5286}.

The $T_{\rm{eff}}$ values of the components derived here are consistent to within 1$\sigma$ with those derived from photometric colours (compare Tables\,\ref{T_atm_par} and \ref{tab:temp}).  The only exception is the $T_{\rm{eff}}$ of the primary component of AL~Ari where this difference is larger than 2$\sigma$.  Fortunately, for AL~Ari we secured three spectra on 2014 December 14, during the total phase of the secondary eclipse when the pure spectrum of the primary component was visible.  By co-adding them we obtained one spectrum with S/N $\sim$ 80.  The use of the {\sl single} mode of {\sl GSSP} gives for this case $T_{\rm{eff}} = 6396 \pm 80$\,K, in 1$\sigma$ agreement with the $T_{\rm{eff}}$ derived from photometric colours. The remaining parameters ($[$M/H$]$, $\xi$, $\zeta^{\star}$, $V_{\rm{rot}} \sin{i}$) are consistent to within 1$\sigma$ with parameters obtained with the {\sl GSSP}$\_${\sl binary} module.

For the systems analysed in this work we obtained generally slightly sub-solar metallicities, with the exception of AL~Ari for which we find a significantly sub-solar value ([M/H] $\sim -0.4$\,dex, see Table\,\ref{T_atm_par}).  In all cases we found that the stars rotate synchronously. In the case of AL~Dor taking the rotational velocity and the macroturbulence simultaneously as free parameters led to unrealistically low values for the former and high values for the latter. Therefore during the fitting process, we fixed the values of $V_{\rm{rot}} \sin{i}$ to the expected values for synchronous rotation (3.7\,km\,s$^{-1}$). Due to the faster rotation of the components of BN~Scl, it was impossible to obtain reliable values of the macroturbulent velocity. {\sl GSSP} returned unrealistically low best-fit values of $\zeta$ with large errors (Table\,\ref{T_atm_par}), therefore we adopted $\zeta = 5$\,km\,s$^{-1}$ based on data from the literature \citep{Gray2001, Sma2014, She2019}.
%mb-finish

\section{Initial Photometric Analysis}

\subsection{Interstellar extinction\label{red}}

We used extinction maps \citep{sch98} with the recalibration by \cite{sch11} to determine the reddening in the direction of all four eclipsing binaries. We followed the procedure described in detail in \cite{such15} assuming the distances from \textit{Gaia} EDR3 \citep{gaia20}. Additionally we used the three-dimensional interstellar extinction map STILISM \citep{cap17}. Finally we adopted the average as the extinction estimate to a particular system.

\subsection{Colour -- temperature calibrations}
\label{temp:col}

To estimate the $T_{\rm eff}$ values of the eclipsing components we collected multi-band magnitudes of the systems. We use 2MASS \citep{cut03} as a good source for infrared photometry and the magnitudes were converted into appropriate photometric systems using transformation equations from \cite{bes88} and \cite{car01}. The reddening (Sect.~\ref{red}) and the mean Galactic interstellar extinction curve from \cite{fit07} assuming $R_V=3.1$ were combined with light ratios from the WD code in order to determine the intrinsic colours of the components. We determined the effective temperaures from a number of colour--temperature calibrations for a few colours: $B\!-\!V$ \citep{alo96,flo96,ram05,gon09,cas10}, $V\!-\!J$, $V\!-\!H$ \citep{ram05,gon09,cas10} and $V\!-\!K$ \citep{alo96,hou00,ram05,mas06,gon09,cas10,wor11}. For few calibrations having the metallicity terms we assumed the metallicity derived from the atmospheric analysis -- see Table~\ref{T_atm_par}. The resulting temperatures were averaged for each component and are reported in Table~\ref{tab:temp}. Usually our colour temperatures are about 1$\sigma$ lower than the temperatures derived from atmospheric analysis in Sec~\ref{temp:atm}.

\begin{table}
\begin{centering}
\caption{The temperatures of components derived from intrinsic colours. }
\label{tab:temp}
\begin{tabular}{lcc}
\hline \hline
System &  \multicolumn{2}{c}{Effective temperature (K)} \\
             &  Primary   &  Secondary \\
\hline
AL~Ari & $6331\pm64$ & $5550\pm50$ \\
AL~Dor & $6032\pm73$ & $6029\pm73$ \\
FM~Leo & $6369\pm50$ & $6356\pm50$ \\
BN~Scl & $6302\pm54$ & $6189\pm53$ \\
\hline
\end{tabular}
\end{centering}
\end{table}

\subsubsection{Adopted values}

Precise determination of the $T_{\rm eff}$ values is very important in our approach because we did not adjust the limb darkening coefficients during a fitting of light curves. Instead, these coefficients are automatically calculated for a given set of surface atmospheric parameters ($T_{\rm eff}$, $\log{g}$) using tables from \cite{VHa93}. Surface gravities are well determined internally within the WD code, however, to set the $T_{\rm eff}$ scale we need external information. The $T_{\rm eff}$  scale was set by fixing the surface $T_{\rm eff}$ of the primary star, $T_1$, to the average of two previous $T_{\rm eff}$ determinations (Sections \ref{temp:atm} and \ref{temp:col}). The adopted $T_1$ in all cases is well within the 1$\sigma$ uncertainty of both $T_{\rm eff}$ determinations. Subsequently the $T_{\rm eff}$ of the secondary, $T_2$, was scaled according to $T_1$ during the light curve analysis with the WD code.

\subsection{Analysis of minima times}
\label{sec:oc}

For the purpose of the analysis we used a collection of minima times from the TIDAK database \citep{kre04}. We augmented the data with minima times determined from the K2 and TESS light curves. Because of the continuous coverage and high photometric precision, the times of minimum derived from K2 and TESS light curves for AL~Dor, FM~Leo and BN~Scl are usually one or even two orders of magnitude more precise than those available in TIDAK.

\subsubsection{AL~Ari}

The case of AL~Ari was discussed by \cite{kim18} who fitted a model with apsidal motion and the light time effect (LITE), caused by an invisible companion, to the minima times. They derived an apsidal period of $U\approx6200$\,yr and found a very eccentric orbit for the third body with an orbital period of $\sim 90$\,yr. We repeated the analysis, but used increased uncertainties for some of the minima times reported. Because of the relatively short time span of observations, instead of fitting a full third-body orbit (five more free parameters) we fitted a model with only a quadratic term (proportional to the first derivative of the sidereal period $P_S$) and the apsidal motion. The computed minima times are:
\begin{equation}
\begin{aligned}
C  &=  T_0 + P_S \cdot E  + b \cdot E^2 +(2j - 3) A_1 \frac{eP}{2\pi} \cos{\omega}\, +\\
& + (j - 1) \frac{P}{2} +  A_2 \frac{e^2P}{4\pi}\sin{2\omega} - (2j-3)A_3\frac{e^3P}{8\pi} \cos{3\omega} %\tau_{\rm AP}(o(e^4))
\end{aligned}
\end{equation}
where $T_0$ is the reference epoch, $E$ is an epoch number, $b=0.5 \dot{P} \bar{P}$, the anomalistic period $P=(2\pi P_S)/(2\pi-\dot{\omega})$, $j=1$ for primary and 2 for secondary eclipses, $\omega$ is the longitude of periastron and $\dot{P}=dP_S/dt$. The mean period $\bar{P}$ was set to 3.747455 d and the orbital eccentricity was set to $e=0.051$. For the coefficients $A_{1,2,3}$ we used equations 16--18 from \cite{gim95}, where we set $\cot^2{i}=0$, because the orbital inclination of AL~Ari is very close to 90$^\circ$.  Altogether we fitted five free parameters: $T_0$, $P$, the linear period change term $b$, $\omega$ at $T_0$ and the rate of advance of the periastron longitude $\dot{\omega}$.

The ephemeris is:
\begin{equation*}
T_{\rm pri}  ({\rm HJD}) = 2451112.8284(4)  +  3.7475431(3) \times E + 1.19(3)\cdot10^{-9} \times E^2 .
\end{equation*}

Fig.~\ref{fig:oc} shows our solution. A significant apsidal motion can be noticed with $\dot{\omega}=(5.7\pm0.2)\times10^{-4}$ deg cycle$^{-1}$ which is a value very close to that reported by \cite{kim18}: $5.9\times10^{-4}$ deg cycle$^{-1}$. The first derivative of the orbital period is quite well constrained to be $\dot{P}=(6.34\pm0.15) \times 10^{-10}$ d cycle$^{-1}$. Interpreting this change in terms of LITE the expected radial velocity acceleration of the barycentre of AL~Ari induced by a third body is $\dot{V}_B=0.19\pm0.01$ m s$^{-1}$ cycle$^{-1}$, which corresponds well with the value derived in Section~\ref{sec:rv} from the radial velocity drift.

\subsubsection{AL~Dor}

This system shows a very slow apsidal motion. To fit the times of minimum we used a model with apsidal motion and a constant orbital period:
\begin{equation}
\begin{aligned}
C &=  T_0 + P_S \cdot E  + (j - 1) \frac{P}{2} + (2j - 3) A_1 \frac{eP}{2\pi} \cos{\omega}\, +\\ 
&+ A_2 \frac{e^2P}{4\pi}\sin{2\omega} - (2j-3)A_3\frac{e^3P}{8\pi} \cos{3\omega} + A_4\frac{e^4P}{16\pi}\sin{4\omega}%\tau_{\rm AP}(o(e^5))
\end{aligned}
\end{equation}
where the coefficients $A$ are given by equations 16--19 in \cite{gim95}. We set the eccentricity to $e=0.195$ and fitted for four free parameters: $T_0$, $P$, $\omega$ at $T_0$, and $\dot{\omega}$.

\begin{figure*}
\begin{minipage}[th]{0.5\linewidth}
\includegraphics[angle=0,scale=0.55]{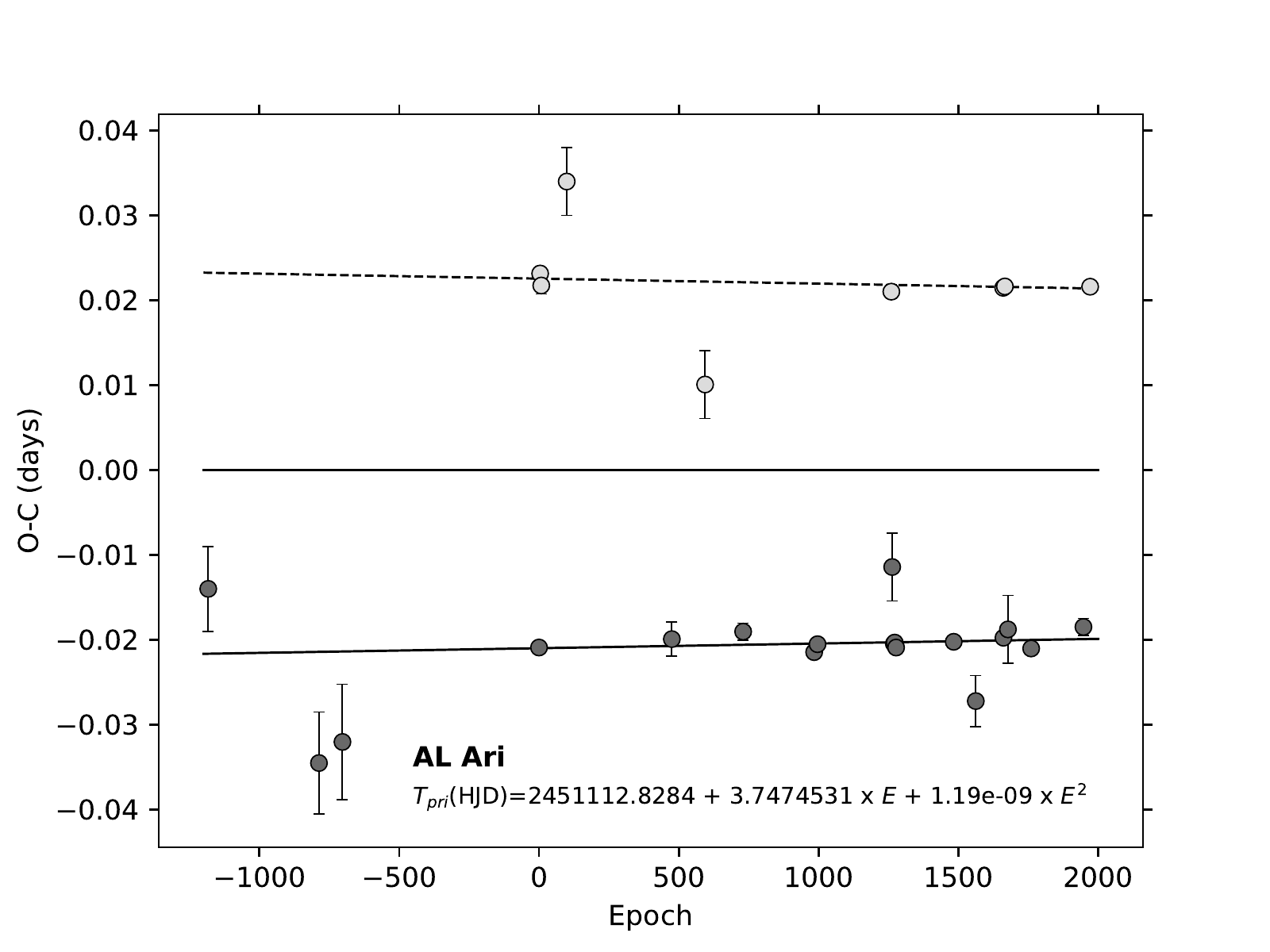} \vspace{-0.15cm}
\mbox{}
\includegraphics[angle=0,scale=.55]{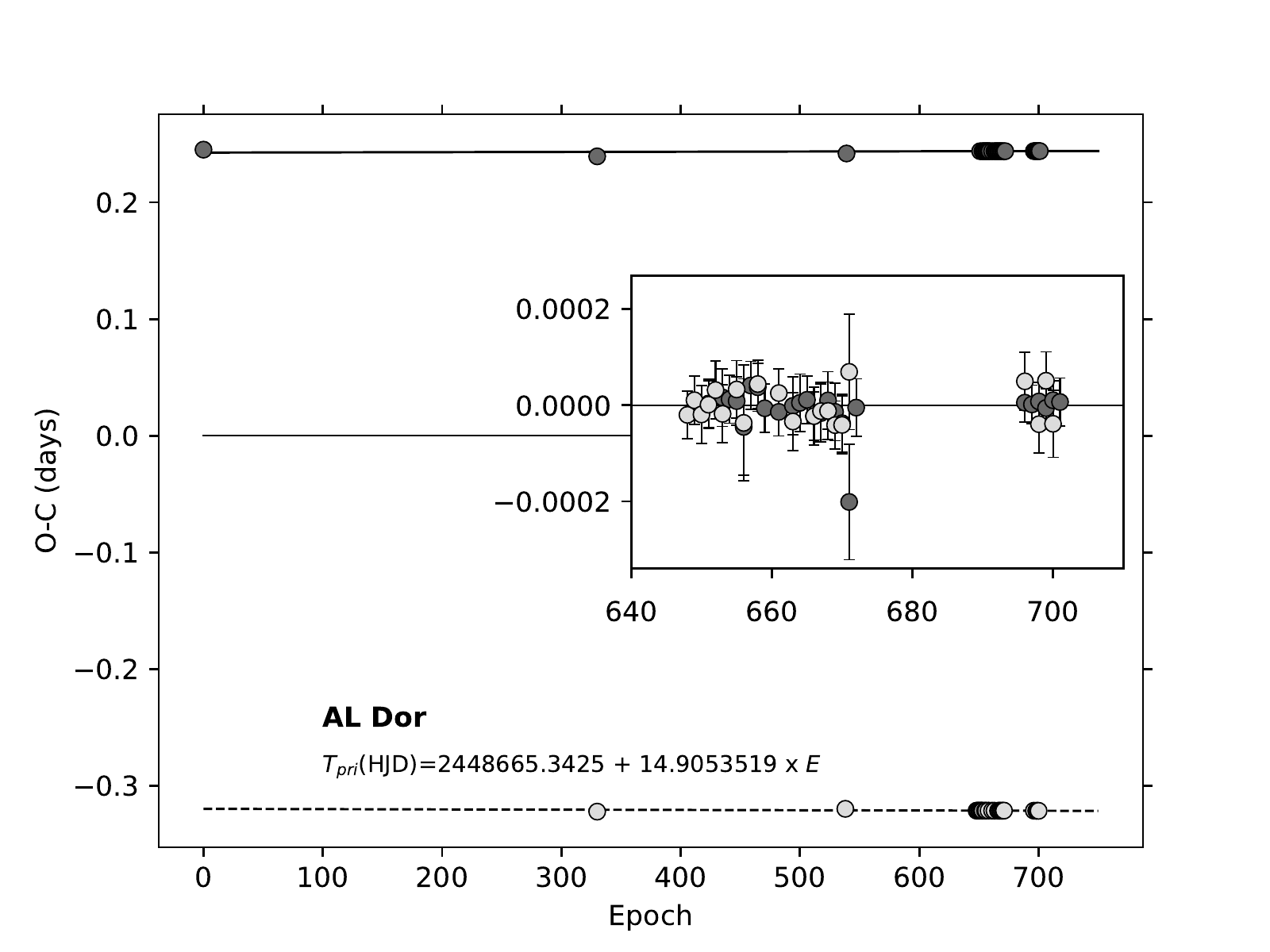}
\end{minipage}\hfill
\begin{minipage}[th]{0.5\linewidth}
\includegraphics[angle=0,scale=.55]{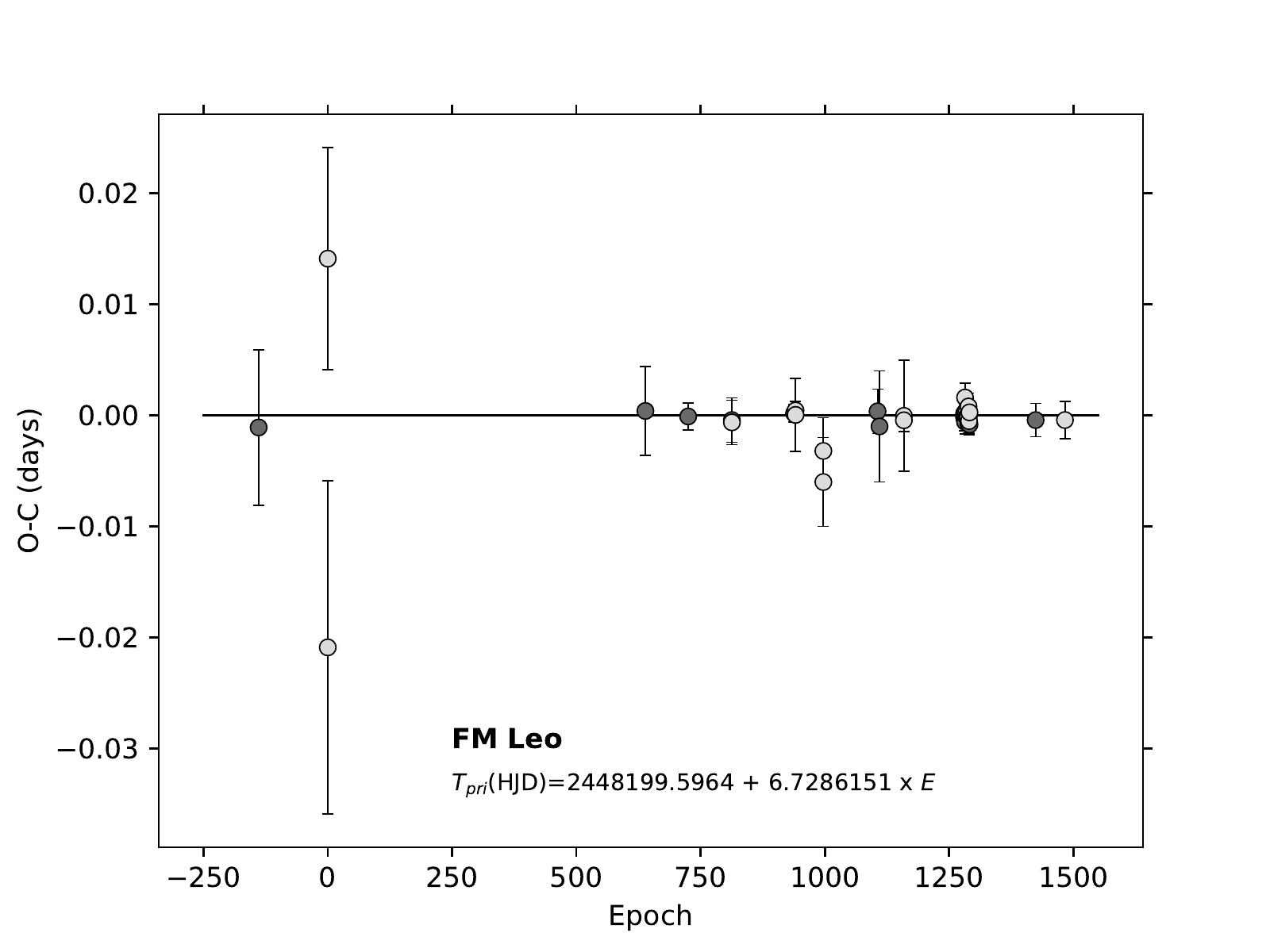}\vspace{-0.15cm}
\mbox{}
\includegraphics[angle=0,scale=.55]{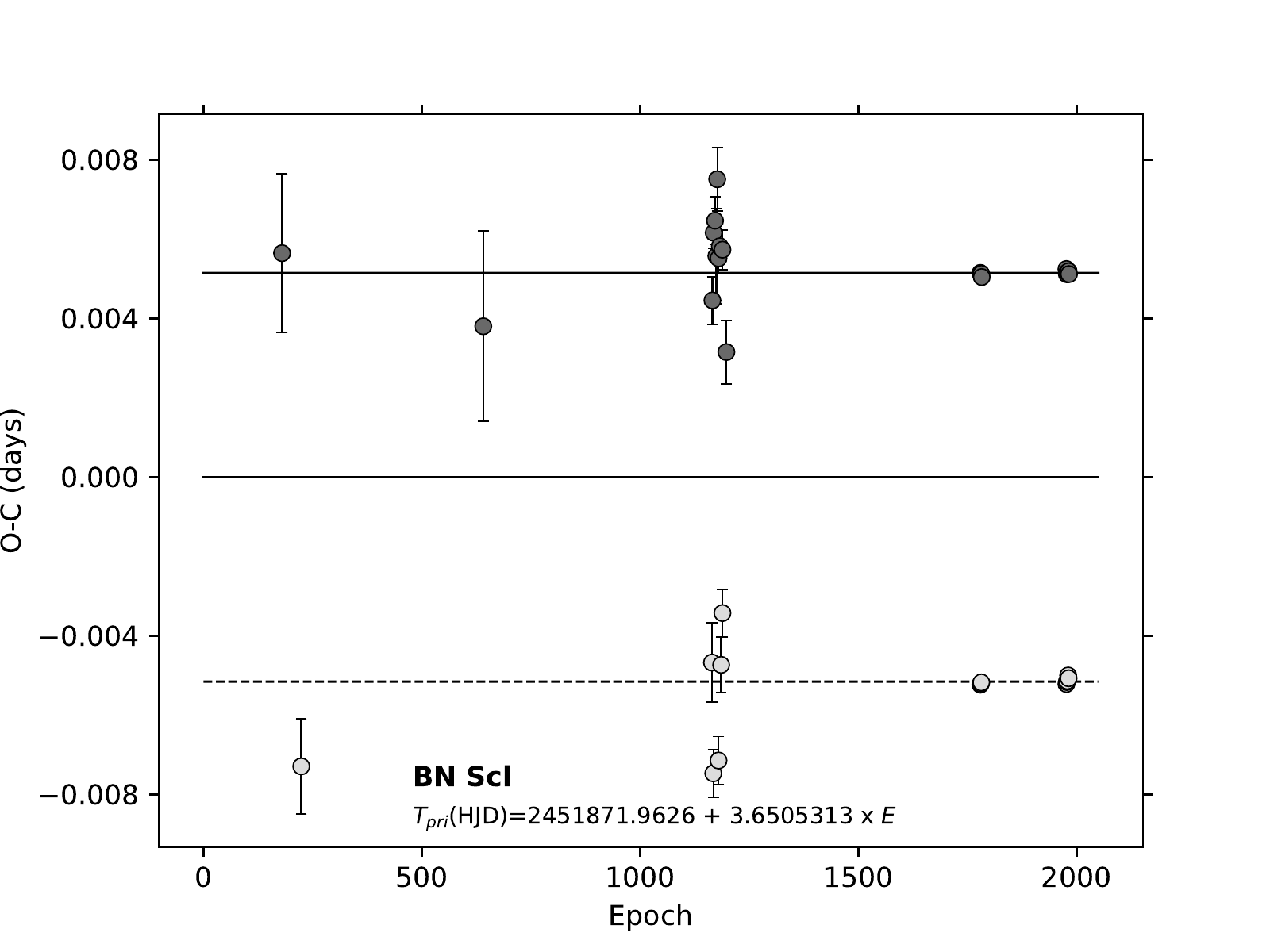}
\end{minipage}\hfill
\caption{$O-C$ diagrams of minima times. A linear term and, in case of AL~Ari also a quadratic term, has been subtracted. Dark grey filled circles correspond to primary minima, and light grey ones to secondary minima. The inset in the panel for AL~Dor shows the residuals of the model fit. \label{fig:oc}}
\end{figure*}

The solution is presented in Fig.~\ref{fig:oc}. The inset shows residuals of our fit to the minima times determined from TESS photometry. Without the apsidal motion term these residuals show a small systematic trend. The resulting ephemeris is:
\begin{equation*}
T_{\rm pri}  ({\rm HJD}) =  2448665.3425(7) + 14.9053519(5) \times E .
\end{equation*}
The rate of periastron advance is $(1.330\pm0.005) \times 10^{-4}$ deg cycle$^{-1}$, corresponding to an apsidal motion period of $U=110\,500\pm400$\,yr.

\subsubsection{FM~Leo}

As the orbit is practically circular we fitted a simple linear ephemeris to both types of minimum:
\begin{equation}
C = T_0 + P_S \cdot E 
\end{equation}
The residuals do not show any noticeable trends (see Fig.~\ref{fig:oc}).
The ephemeris for FM~Leo is:
\begin{equation*}
T_{\rm pri}  ({\rm HJD}) = 2448199.5964(14)  +  6.7286151(12) \times E . 
\end{equation*}

\subsubsection{BN~Scl}

The orbit of the system is slightly eccentric and we fitted a linear ephemeris with no apsidal motion or LITE terms:
\begin{equation}
C =  T_0 + P_S \cdot E  + (j - 1) \frac{P}{2} + (2j - 3) A \frac{eP}{2\pi} \cos{\omega}
\end{equation}
where the coefficient A is given by equation 16 in \cite{gim95}. The eccentricity and longitude of periastron were set to values from the full analysis of the light and velocity curves (Section~\ref{full:bnscl}). The resulting ephemeris is:
\begin{equation*}
T_{\rm pri}  ({\rm HJD}) = 2451871.963(3)  +  3.6505313(2) \times E .
\end{equation*}

\section{Analysis of combined light and radial velocity curves \label{wd}}

For analysis of the eclipsing binaries we make use of the Wilson-Devinney program version 2007 \citep{wil79,wil90,van07}\footnote{\texttt{ftp://ftp.astro.ufl.edu/pub/wilson/lcdc2007/}}, equipped with a Python wrapper.

\subsection{Initial parameters}

We fixed the $T_{\rm eff}$ of the primary component during analysis to the average of the $T_{\rm eff}$ values derived from the colour-temperature calibrations and the atmospheric analysis. In all cases those two determinations are consistent to within 1$\sigma$. The standard albedo and gravity brightening coefficients for convective stellar atmospheres were chosen. The stellar atmosphere option was used (\verb"IFAT1=IFAT2=1"), radial velocity tidal corrections were automatically applied (\verb"ICOR1=ICOR2=1") and no flux level dependent weighting was used. We assumed synchronous rotation for both components in all systems. The epoch of the primary minimum was set according to results of the $O-C$ diagram analysis. Both the logarithmic \citep{kli70} and square root  \citep{dia92} limb-darkening laws were used, with coefficients tabulated by \cite{VHa93}.

\subsection{Fitting model parameters}

With the WD binary star model we fitted simultaneously the available light curves and a radial velocity curve for each component. We assumed a detached configuration in all models and a simple reflection treatment (\verb"MREF = 1", \verb"NREF = 1"). Each observable curve was weighted only by its {\it rms} through comparison with the calculated model curve. We adjusted the following parameters during analysis: the orbital period $P_{\rm orb}$, the semimajor axis $a$, the mass ratio $q$, both systemic radial velocities $\gamma_{1,2}$, the phase shift $\phi$, the eccentricity $e$, the argument of periastron $\omega$, the orbital inclination $i$, the temperature of the secondary $T_2$, the modified Roche potentials $\Omega_{1,2}$ -- corresponding to the fractional radii $r_{1,2}$ -- and the luminosity parameter $L_1$. Additionally, we fitted for third light $l_3$. %THIS WAS ALREADY SAID IN THE PREVIOUS SECTIONL We used stellar atmosphere approximation in both stars \verb"IFAT1=IFAT2=1" and proximity effects were included in models: \verb"ICOR1=ICOR2=1".
The best models were chosen according to their reduced $\chi^2$ and a lack of significant systematic trends in the residuals.

The statistical (formal) errors on model parameters were estimated with the Differential Correction subroutine of the WD code. We added to the formal errors a mean correction on every model parameter from the last three iterations, provided that the iterated models were close to the final solution (to within $1\sigma$). We also compared numerical values of model parameters obtained with different limb darkening laws and with a third light accounted for. Averaged differences were added in quadrature to the total error budget.

\begin{figure*}
\begin{minipage}[th]{0.5\linewidth}
\includegraphics[angle=0,scale=0.5]{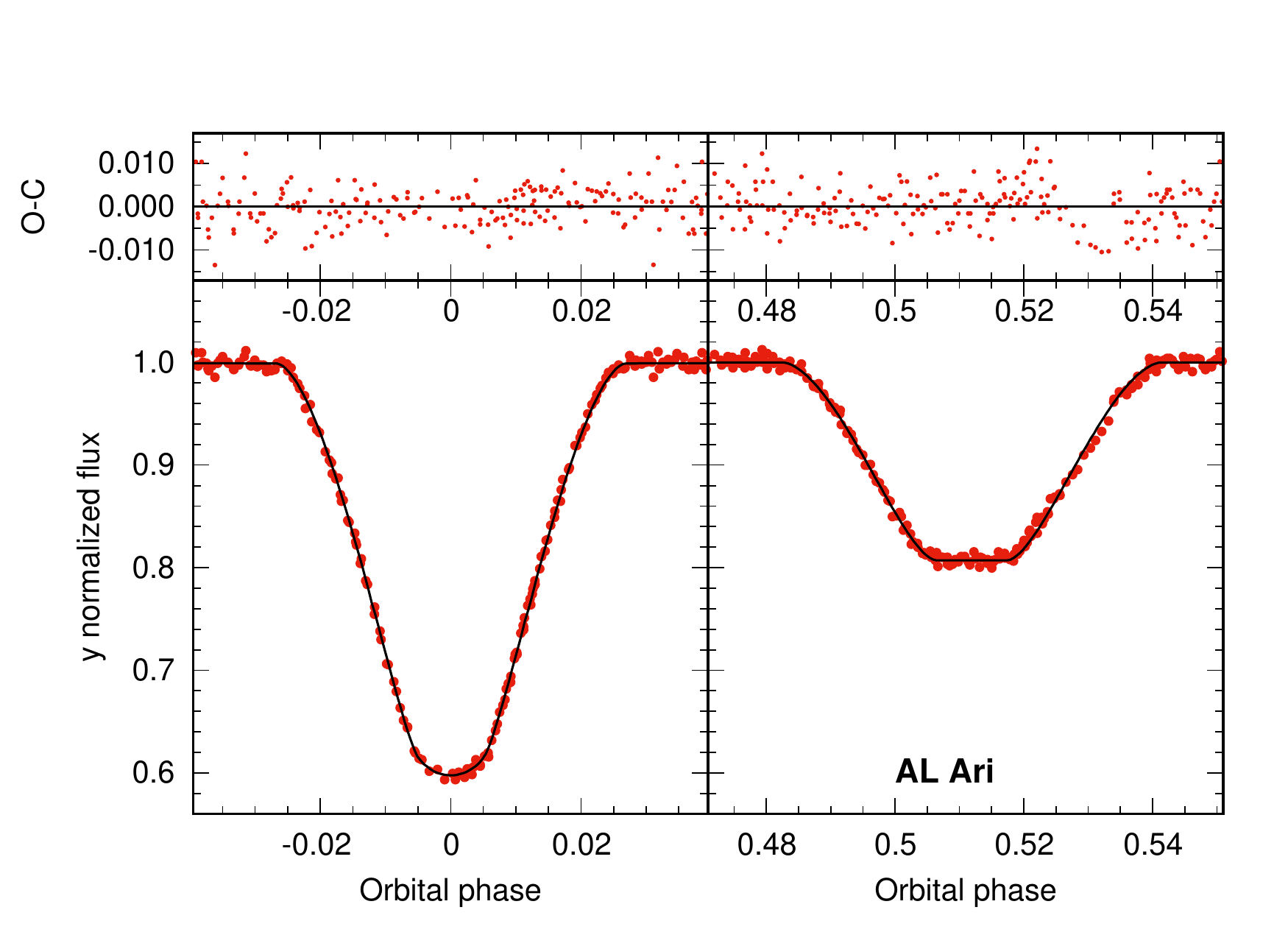} \vspace{-1.0cm}
\mbox{}
\includegraphics[angle=0,scale=.5]{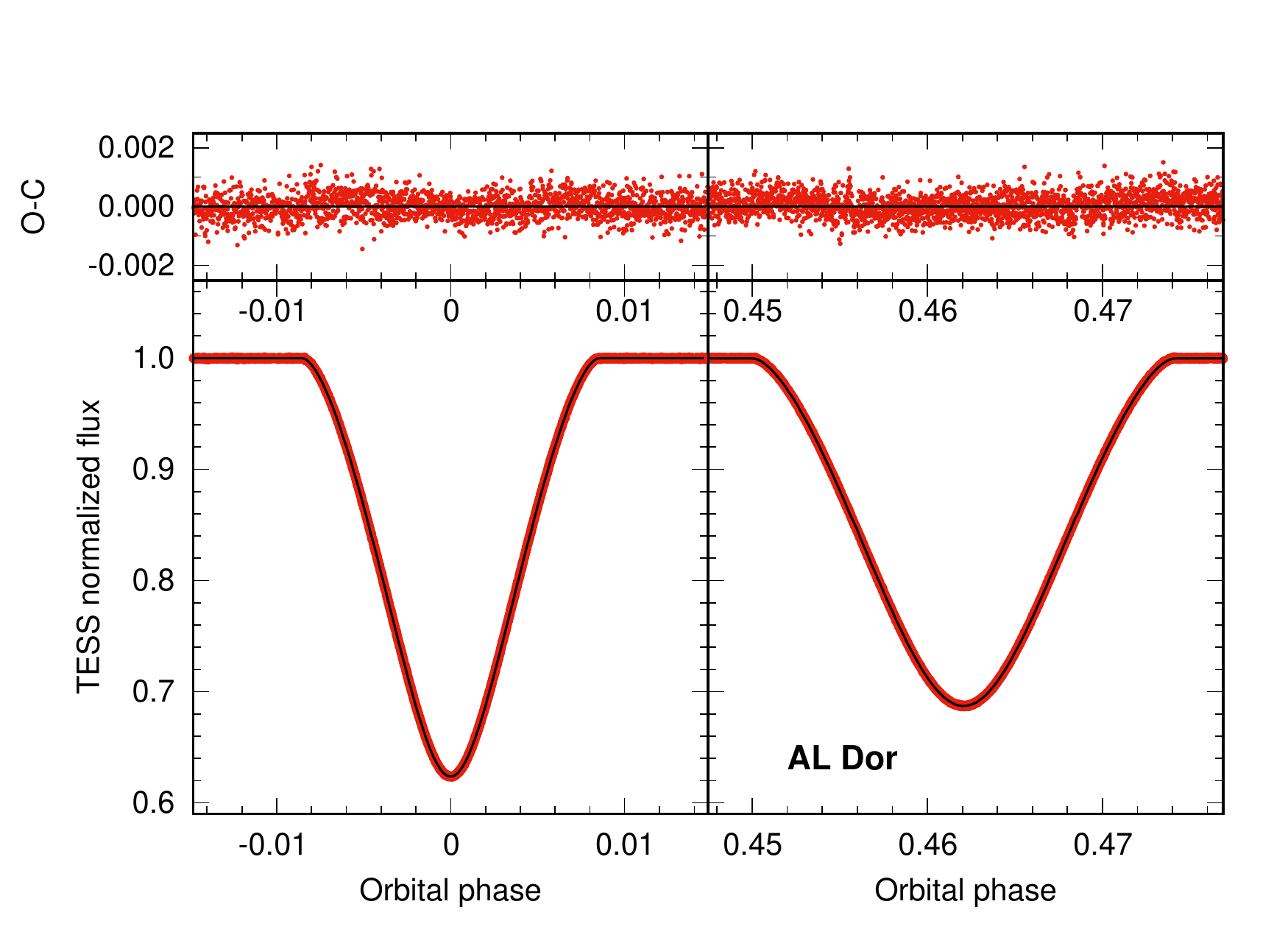}
\end{minipage}\hfill
\begin{minipage}[th]{0.5\linewidth}
\includegraphics[angle=0,scale=.5]{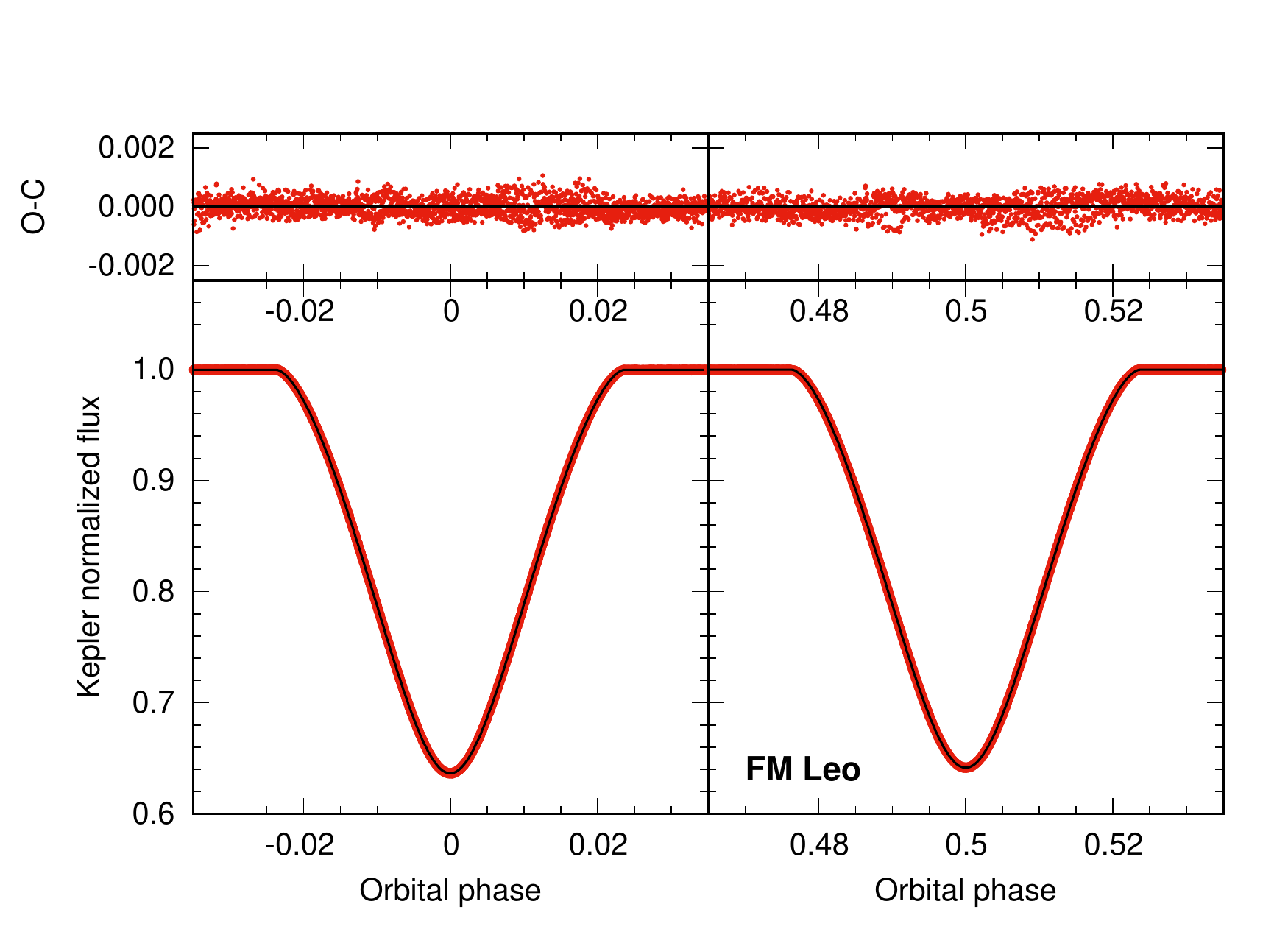}\vspace{-1.0cm}
\mbox{}
\includegraphics[angle=0,scale=.5]{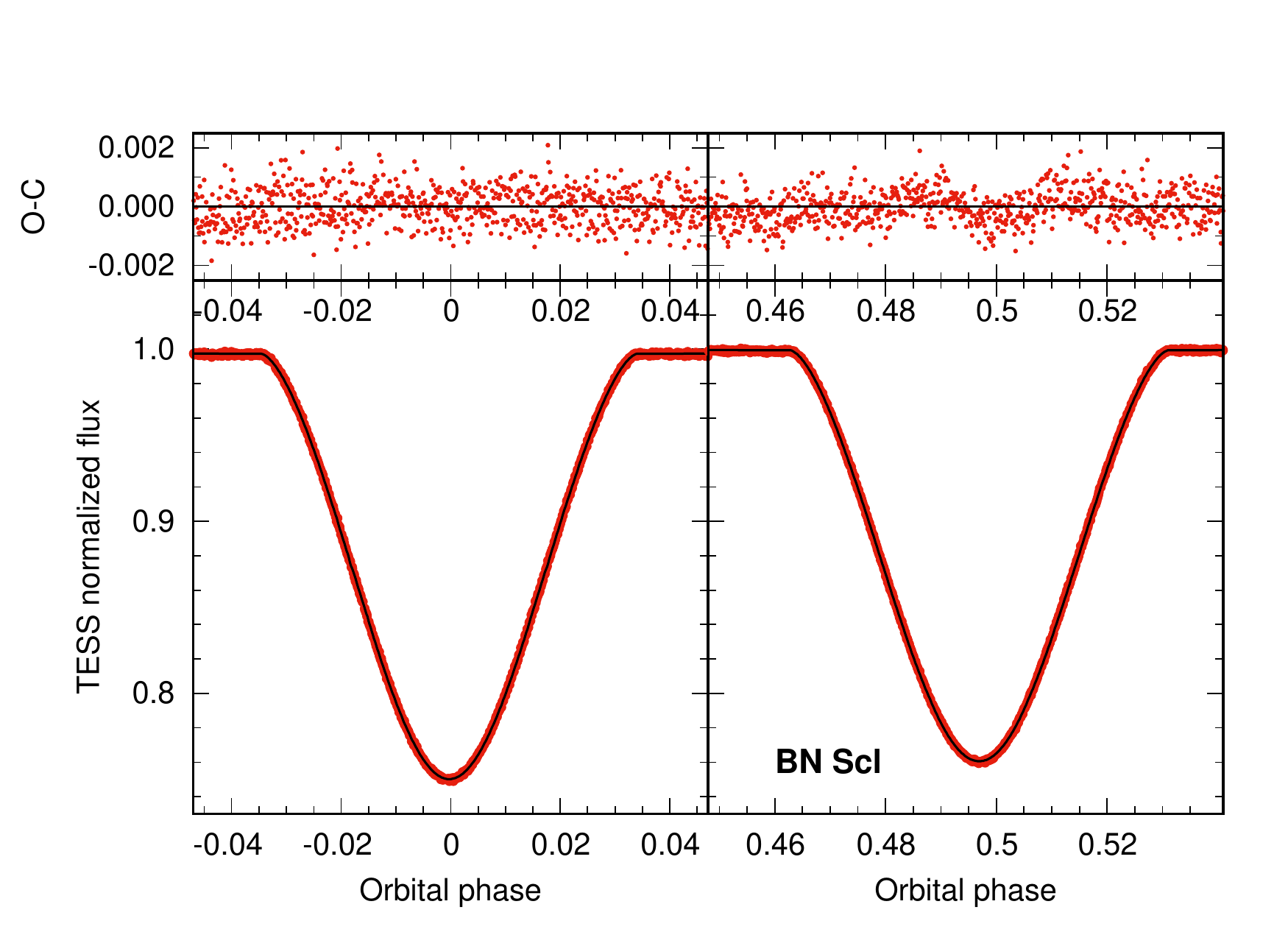}
\end{minipage}\hfill
\caption{The WD model fits to the photometric observations. \label{fig:light}}
\end{figure*}

\begin{figure*}
\begin{minipage}[th]{0.5\linewidth}
\includegraphics[angle=0,scale=0.5]{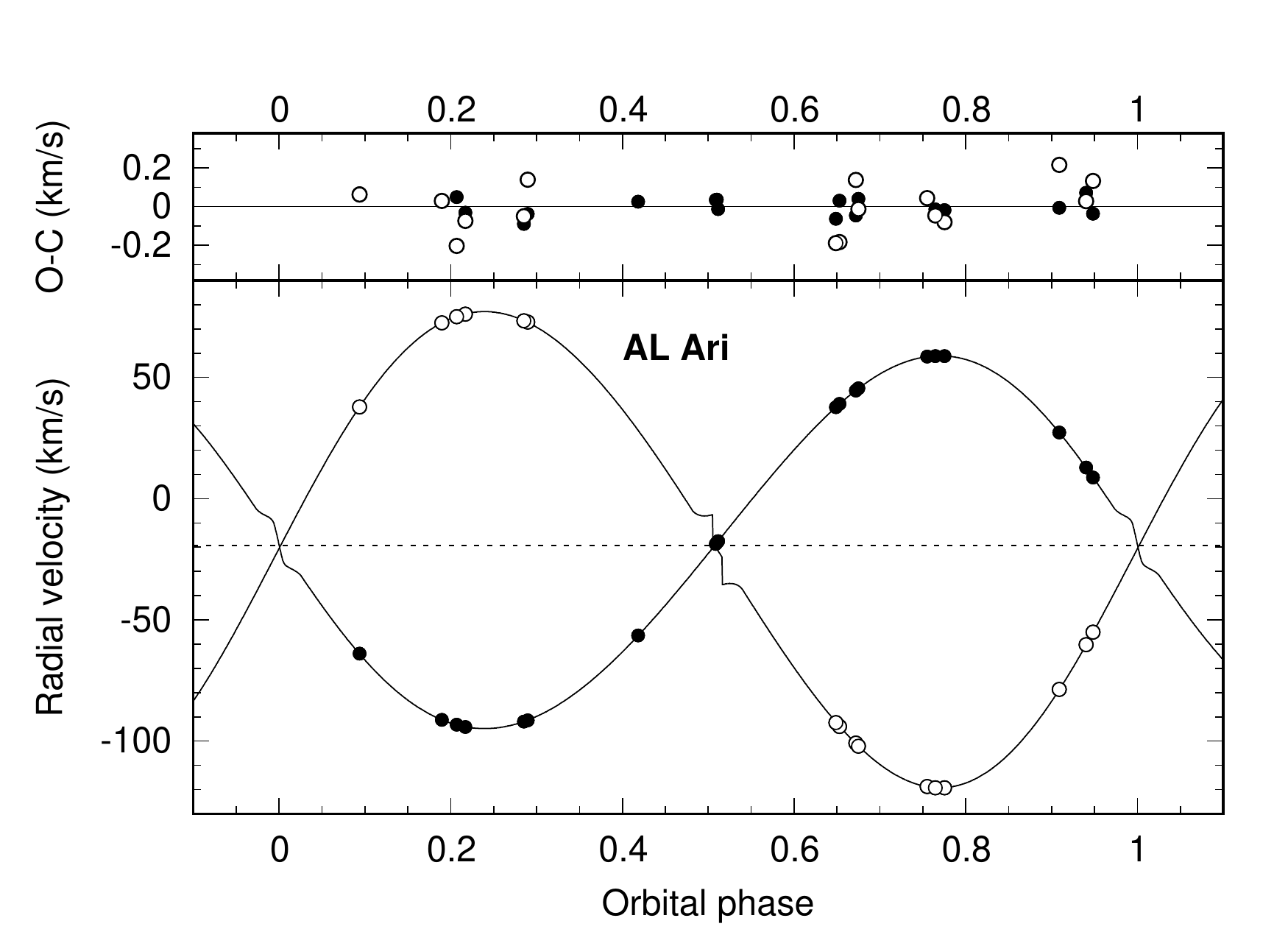} \vspace{-1.0cm}
\mbox{}
\includegraphics[angle=0,scale=.5]{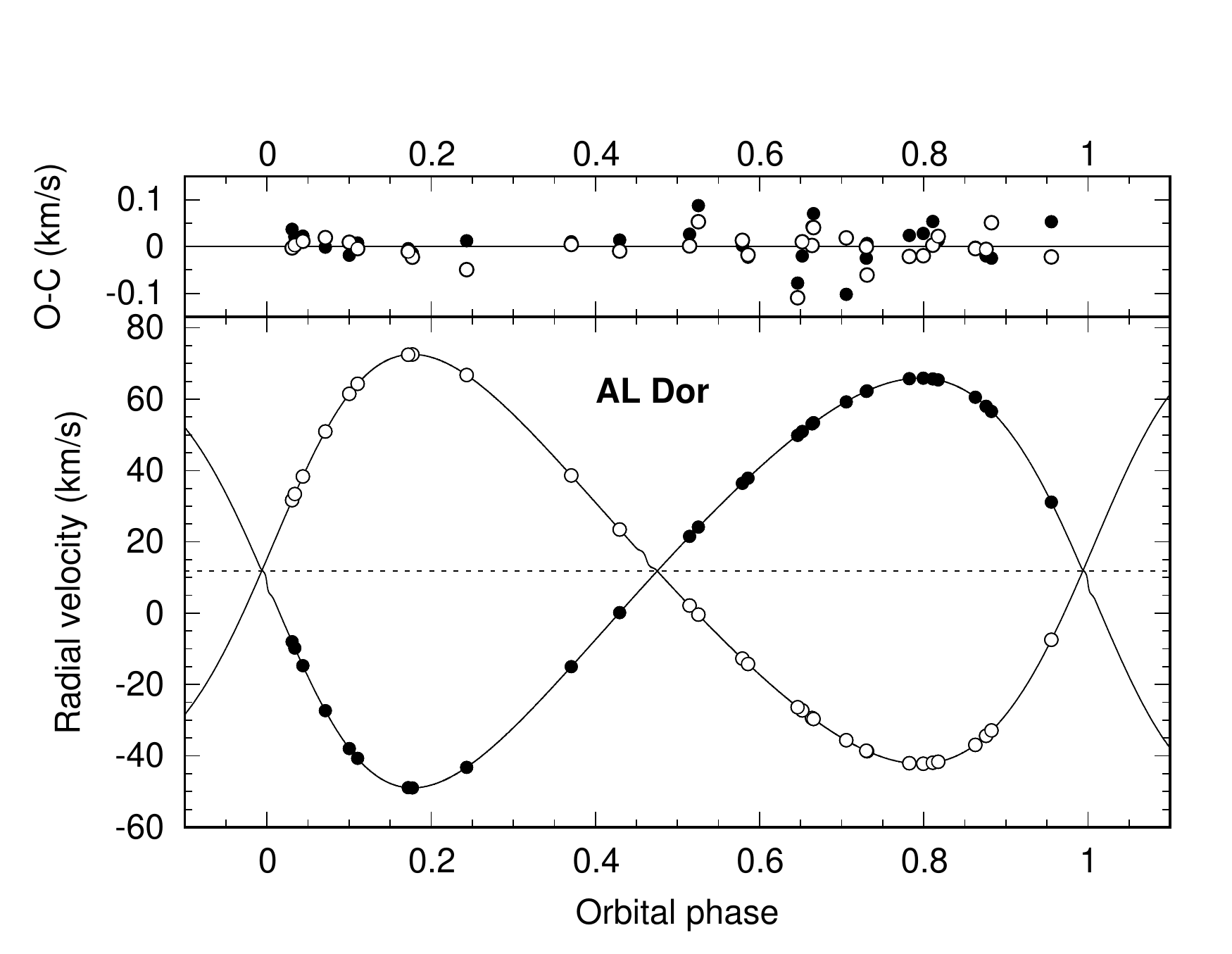}
\end{minipage}\hfill
\begin{minipage}[th]{0.5\linewidth}
\includegraphics[angle=0,scale=.5]{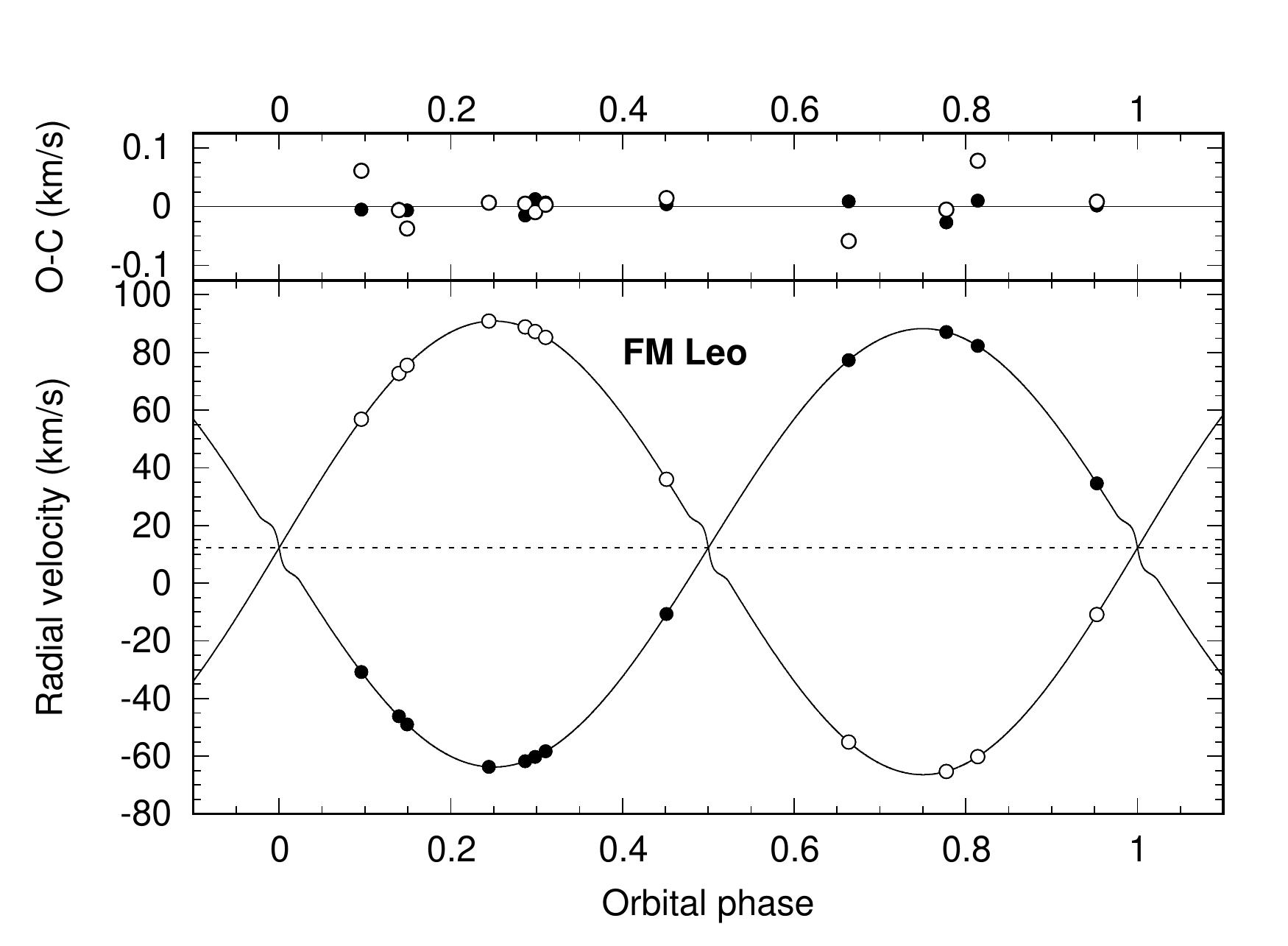}\vspace{-1.0cm}
\mbox{}
\includegraphics[angle=0,scale=.5]{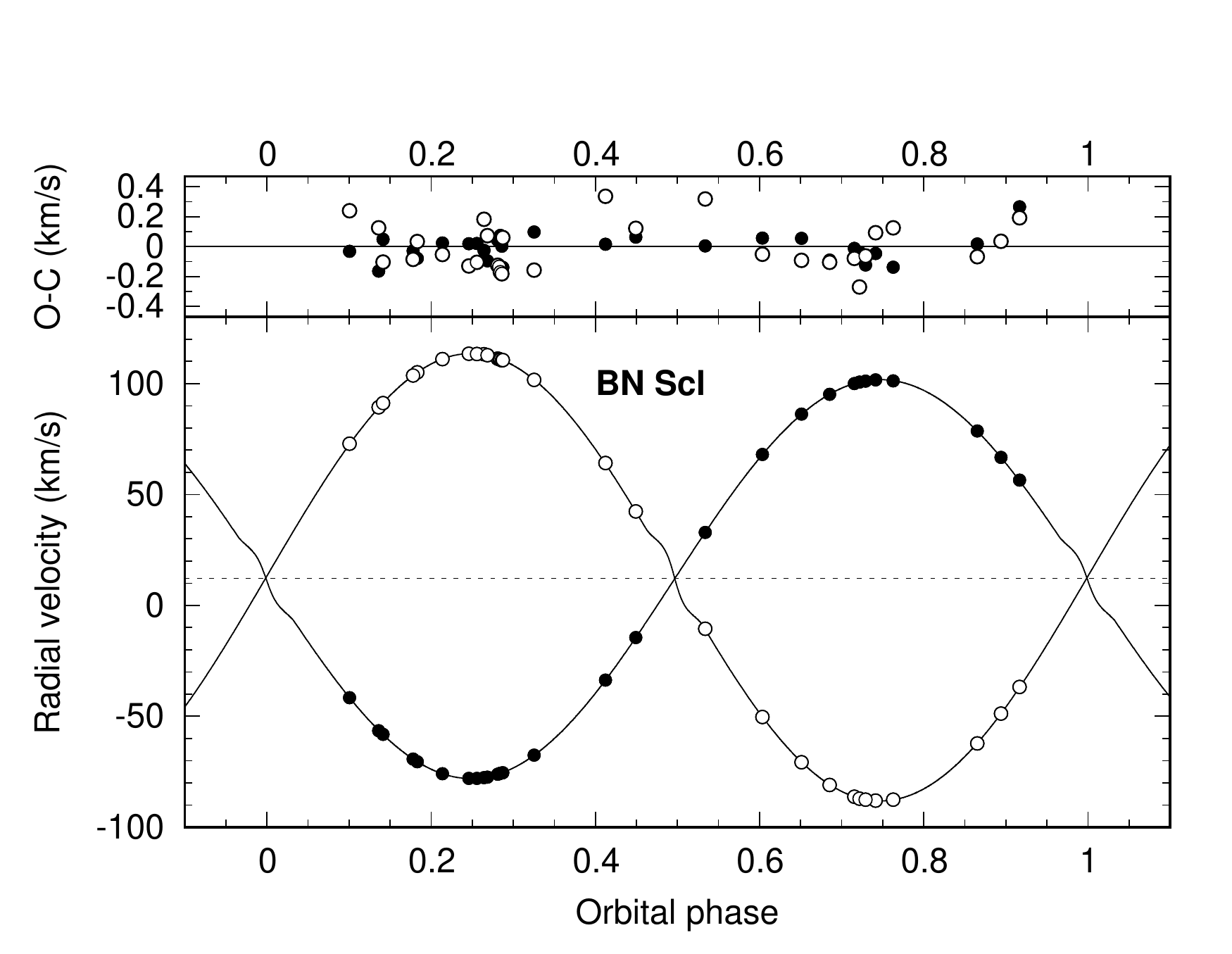}
\end{minipage}\hfill
\caption{The WD model fits to the radial velocity curves. \label{fig:rv}}
\end{figure*}

\subsection{Analysis details and results}
\label{WD_results}

\subsubsection{AL~Ari}

We used three Str{\"o}mgren bands ($v$, $b$ and $y$) because the light curve in the $u$ filter is significantly more noisy and also has additional systematic offsets in the secondary minimum due to unfavourable sky conditions. The grid fineness parameters were set to \verb"N1=N2=50". In the beginning we solved the radial velocities and light curves separately to obtain the best individual solutions. We then combined the light curves with the velocimetry to obtain a simultaneous solution with the WD code. As AL~Ari has a notable apsidal motion and acceleration of its barycentre due to an invisible companion, we included two additional free parameters in the fit: the first derivative of the orbital period \verb"DPDT"$\,=\!\dot{P}$ (dimensionless) and the first derivative of the longitude of periastron \verb"DPERDT"$\,=\!\dot{\omega}$ (rad day$^{-1}$). We set the initial epoch at the mid-time around which the Str{\"o}mgren light curves were obtained: May 1998. Because there were many correlations between parameters in the beginning we set \verb"DPERDT"$\,=\!2.7\times10^{-5}$ based on results from the analysis of minima times. After several iterations we included this as a free parameter.

The final solution is consistent with results from the minima times analysis, however the resulting orbital period change is five times faster. It is probable that more complicated period changes (caused by a tertiary) beyond the simple linear model we used are responsible for this. We tested also for the presence of third light. However, its inclusion did not improve the solution and the fitted values of $l_3$ were consistent with zero in all bands. We could only put an upper limit of $l_3<1$\% for the light curves. The parameters of the final solution are given in Table~\ref{tab_par_orb}. The best model fit to the $y$-band light curve is presented in Fig.~\ref{fig:light}, while the best model fit to the radial velocities is shown in Fig.~\ref{fig:rv}. The radial velocity {\it rms} of the primary component is consistent with expected precision of HARPS in EGGS mode, however the secondary shows significantly larger {\it rms} than expected.

The configuration of the system is well detached. The tidal deformations, defined as $(r_{\rm point}-r_{\rm pole})/r_{\rm mean}$, are just 0.3\% and 0.1\% for the primary and the secondary, respectively.  The absolute dimensions of the system were calculated using nominal astrophysical constants advocated by IAU 2015 Resolution B3 \citep{prsa16} and are presented in Table~\ref{par_fi}. Unlike the other systems analysed in this work, the components of AL~Ari have quite different masses, sizes, temperatures and luminosities. 
%Because the masses of the components straddle that of the Sun, the system is interesting for testing convective core overshooting at lower metallicities.

We estimated the spectral type of the system to be F6\,V + G7\,V using tables by \cite{pec13}. This is consistent with the spectral type of F8 estimated from photographic observations \citep{hil52}.

\subsubsection{AL~Dor}

We combined the TESS light curve with our velocimetry from HARPS and CORALIE spectra. We used the grid fineness parameters \verb"N1=N2=60", the highest available value in the 2007 version of the WD code. To represent the TESS band we used the Cousins $I$-band (filter 16 in the WD code). Our analysis of the TESS photometry of the AI~Phe system \citep{max20} showed that approximating the TESS band with the $I_C$ band leads to systematics in the fractional radii which are smaller than 0.05\% and thus can be neglected.

The logarithmic law of the limb darkening was applied to both stars. Because of the apsidal motion there is a small orbital phase shift between the velocimetry and photometry. To achieve the best solution we first solved only the velocimetry to derive the orbital parameters, and then used them as input to derive a photometric model using only the photometry. We iterated this step several times until all parameters of the two models were the same, with the exception of $\phi$ and $\omega$. Then we ran several simultaneous iterations setting \verb"DPERDT" free. The resulting value of $\dot\omega = (1.2\pm0.1) \times 10^{-4}$ deg cycle$^{-1}$ is in full agreement with the result from the minima times analysis. A summary of the model parameters is given in Table~\ref{tab_par_orb}. Adjusting $l_3$ does not improve the solution, as the WD code returns small negative values of $l_3$ without a decrease of residuals. We assumed $l_3=0$.

The components of AL~Dor are extremely similar to each other (Table~\ref{par_fi}): their masses and temperatures are the same to within 0.1\% and the radii to within 0.6\%. Because of this the system is valuable for very precise calibration of the surface brightness -- colour relations: the components have the same colours and the flux ratio does not change significantly for a wide range of wavelengths.

We estimated the spectral type of the system as F9\,V + F9\,V based on the calibration by \cite{pec13}, which is consistent with the spectral type of F8\,V given by \cite{hou75}. Our mass measurements coincide well with the measurements from \cite{gal19} because we used the same velocimetry, however, the errors on our masses are larger by a factor of 1.5. Note that we use a different definition of which is the primary star, because the eclipse of the less massive and smaller star is slightly deeper than the other eclipse.

%Components are very similar each other with the secondary being slighly

\begin{table*}
\begin{centering}
\caption{Results of the final analysis with the WD code.}
\label{tab_par_orb}
\begin{tabular}{lcccc}
\hline \hline
Parameter & AL~Ari & AL~Dor & FM~Leo & BN~Scl\\
\hline
\multicolumn{5}{c}{Orbital parameters} \\
$P$ (d) & 3.7474513(8)&14.9053530(12)&6.7286133(8)& 3.65053206(14) \\
T$_0$ fixed (HJD) & 2450951.66638&2458368.7278&2452499.18235& 2452598.4254 \\
$a \sin{i}\, (R_\odot)$ & 12.9492(40)&33.1676(64)&20.5706(22)& 13.7675(61)  \\
$q=M_2/M_1$ & 0.7828(5)&1.0009(6)&0.9665(2)&  0.8914(8)  \\
$\gamma_1$ (km~s$^{-1}$) & $-$19.31(4) & 11.806(6) &12.220(7) &  12.19(4)\\
$\gamma_2$ (km~s$^{-1}$) & $-$18.95(4) & 11.810(6) &12.256(7) & 12.27(4)\\
$e$ & 0.0507(9)&0.1950(2)&0.00013(3)&0.0051(2) \\
$\omega$ (deg) & 68.5(3)&107.47(2)&67(14)& 209(5) \\
$\dot{\omega}$ (deg year$^{-1}$) &0.075(17)&$0.0029(3)$&--&-- \\
$\dot{P}$ (d year$^{-1}$) & 3.8$\cdot10^{-7}$&-- &-- &-- \\
&&&&\\
\multicolumn{5}{c}{Photometric parameters} \\
$\Delta \phi$ (phase shift)& 0.00591(3) & $-0.016309(3)$ &0.000061(4) &$-$0.00228(8)\\
$ i$ (deg) & 89.30(8)&88.993(10)&87.941(23)& 84.65(8) \\
$T_2/T_1$ & 0.8751(9)&0.9995(2)&0.9983(2)&0.9837(10)\\
$\Omega_1$ &10.273(60)&31.629(55) &13.643(15)&8.310(52)\\
$\Omega_2$ &12.373(98)&31.471(54)&14.212(24)&9.997(95)\\
$r_1$ &0.10592(67)&0.03291(6)&0.07892(10)&0.13268(53)\\
$r_2$ &0.06985(61)&0.03311(6)&0.07327(14)&0.09983(64)\\
$T_{\rm1,eff}$ fixed (K) &6363&6056&6397& 6305\\
$A_1$ &0.5(fixed)&0.5(fixed)&0.20(1)&0.55(2)\\
$A_2$ &0.5(fixed)&0.5(fixed)&0.23(1)&0.00(6)\\
$L_2/L_1(v)$&0.1768(9)&--&--&--\\
$L_2/L_1(b)$ &0.2146(7)&--&--&--\\
$L_2/L_1(y)$ &0.2366(7)&--&--&--\\
%$L_2/L_1$(WASP) &--&--&--&0.4808(11)\\
$L_2/L_1$({\it Kepler}) &--&--&0.8562(10)&--\\
$L_2/L_1$(TESS) &--&1.0106(10)&--&0.5376(11)\\
$l_3$ & 0.00(1)& 0 & 0.0050(4)& 0.000(1)\\
&&&&\\
\multicolumn{5}{c}{Derived quantities} \\
$L_2/L_1(V)$ &0.2355&1.0100&0.8550&0.5250\\
$L_2/L_1(K)$ &0.3715&1.0116&0.8606&0.5563\\
$a (R_\odot)$ & 12.9503(41)&33.1727(65)&20.5839(22)&13.8252(63)   \\
$K_1$ (km~s$^{-1}$) & 76.863(20) & 57.418(17) &76.017(10)&89.907(42)\\
$K_2$ (km~s$^{-1}$) &  98.186(51)& 57.365(14) &78.654(15)&100.863(72)\\
RV$_1$ $rms$ (m~s$^{-1}$) & 46 &38 &11& 96 \\
RV$_2$ $rms$ (m~s$^{-1}$) & 124 &29 &37& 152\\
$v$ $rms$ (mmag) &5.05&--&--&--\\
$b$ $rms$ (mmag) &5.04&--&--&--\\
$y$ $rms$ (mmag) &5.03&--&--&-- \\
%WASP $rms$ (mmag) &--&--&--&9.44\\
{\it Kepler} K2 $rms$ (mmag) &--&--&0.27&--\\
TESS $rms$ (mmag) &--&0.42&--&0.74\\
%$\chi^2/DOF$  & \multicolumn{2}{c}{?}  \\
\hline
\end{tabular}
\end{centering}
\end{table*}

\begin{table*}
\centering
\caption{Physical parameters of the stars.}
\label{par_fi}
\begin{tabular}{@{}lcccccccc@{}}
\hline \hline
 & \multicolumn{2}{c}{AL~Ari} &  \multicolumn{2}{c}{AL~Dor}&  \multicolumn{2}{c}{FM~Leo}& \multicolumn{2}{c}{BN~Scl} \\\hline
Parameter& Primary& Secondary & Primary& Secondary &Primary& Secondary&Primary& Secondary\\
\hline
$M$ ($M_{\sun}$) & 1.1640(13)& 0.9112(7)&1.1018(6)&1.1028(7)&1.3144(5)&1.2703(4)&1.4067(22)&1.2539(15)\\
$R$ ($R_{\sun}$) & 1.372(9)& 0.905(8)&1.0917(20)&1.0983(20)&1.6245(21)&1.5082(29)&1.834(7)&1.380(9)\\
$\log{g}$ (cgs) & 4.229(5) & 4.485(8) &4.404(2)&4.399(2)&4.135(1)&4.185(1)&4.059(3)&4.256(6)\\
$T_{\rm eff}$ (K) & 6363(64)& 5560(67)&6056(60)&6050(60)&6397(56)&6386(57)&6305(55)&6236(74)\\
$L$ ($L_{\sun}$) &2.78(12)& 0.71(4)&1.44(6)&1.46(6)&3.98(14)&3.41(12)&4.79(17)& 2.59(13)\\
$\upsilon_{\rm syn}$ (km s$^{-1}$)&  18.53(12) & 12.22(11) &3.71(1)&3.73(1)&12.22(2)&11.34(2)&25.43(10)&19.13(12) \\
%$\upsilon_{\rm t}$ (km s$^{-1}$) &1.73(13) & 2.0(5)& 1.36(18)&1.29(17)&1.61(15)&1.49(17)&1.62(17)&1.7(3) \\
%$\upsilon_{\rm macro}$ (km s$^{-1})\,^b$ &4.7(3.3) & 2.6(2.0) &3.1(5)&3.4(5)&5.8(1.2)&4.3(1.5)&{\it 5}&{\it 5} \\
$[{\rm M}/{\rm H}]$ (dex) &\multicolumn{2}{c}{$-0.44(8)$} &\multicolumn{2}{c}{$-0.11(5)$}&\multicolumn{2}{c}{$-0.10(5)$}&\multicolumn{2}{c}{$-0.20(6)$} \\\hline
%Age (Gyr)& \multicolumn{2}{c}{$\div$}&\multicolumn{2}{c}{--}&\multicolumn{2}{c}{--}&\multicolumn{2}{c}{--}\\
$D_{\rm phot}$ (pc) & \multicolumn{2}{c}{140(2)} &\multicolumn{2}{c}{66(1)}&\multicolumn{2}{c}{144(2)}& \multicolumn{2}{c}{184(3)}\\
$\varpi_{\rm phot}$ (mas) &\multicolumn{2}{c}{7.13(11)} & \multicolumn{2}{c}{15.13(20)}&\multicolumn{2}{c}{6.91(10)} &\multicolumn{2}{c}{5.44(8)}\\
$E(B\!-\!V)$ (mag) & \multicolumn{2}{c}{0.017(10)}&\multicolumn{2}{c}{0.002(2)}&\multicolumn{2}{c}{0.018(10)}&\multicolumn{2}{c}{0.006(4)}\\
\hline
\end{tabular}
%\\$^a$ From $T_{\rm eff}$ using Pecaut \& Mamajek (2013) calibration.
%\\$^b$ From Gray (2005) calibration for main sequence stars.
\end{table*}

\subsubsection{FM~Leo}
\label{full:fmleo}

We fitted the light and radial velocity curves simultaneously, using the grid fineness parameters \verb"N1=N2"=60. To represent the {\it Kepler} band within the WD code we used the Cousins $R$ band (filter 15 in the WD code) which has very similar effective wavelength. For our initial solution we used parameters from \cite{rat10}. We quickly found that a circular orbit produces small systematic trends in the residuals during both eclipses, so we set the eccentricity and the longitude of periastron as free parameters during the fitting. Initially we set the third light to zero and looked for the best solution with the logarithmic law of the limb darkening (parameter \verb"LD" $=$ $-2$ for both components). In order to properly recreate the out-of-eclipse light changes, which have an amplitude of 0.6 mmag, we also fitted the albedo parameters of both stars, $A_1$ and $A_2$.  The best light curve solution had a very small {\it rms} of 0.283 mmag, but there were still small systematics noticable in the eclipses at a level of $\sim 300$ ppm. We tried also the square root law of the limb darkening (\verb"LD" $=$ $-3$) and obtained a slightly better solution with an {\it rms} of 0.275 mmag but the systematic residuals persisted, albeit somewhat smaller ($\sim 200$ ppm). Finally, we let third light be a free parameter of the fit. Surprisingly a small third light $l_3=0.005$ was detected with a high significance of 12$\sigma$. The best solution with $l_3$ has no systematic residuals within the primary minimum and within the secondary minimum they are of order of only $\sim 100$ ppm (see the upper right panel of Fig.~\ref{fig:light}). The model parameters are given in Table~\ref{tab_par_orb}. We note also an anomalous albedo for both components of FM Leo, which is expected to be 0.5 in case of a convective atmosphere, but it is significantly lower. The reason of that is unclear.

It is interesting that the best orbital solution we found still has a very small eccentricity. It is not an artefact or a spurious detection, because only a non-circular orbit removes the systematic residuals in eclipses. Assuming that FM~Leo is a triple system (see discussion below), the tidal forces of the tertiary would induce perturbations to the inner orbit of FM~Leo, producing tiny but noticeable deviations from circularity.

The components, as in the case of AL~Dor, have very similar temperatures, however they differ slightly in mass and radius (see Table~\ref{par_fi}). \cite{max18} already pointed out that this system can be very valuable for testing stellar models, because both components are a bit evolved and they have masses in a range where the core-overshooting starts to play an important role \citep[e.g.][]{cla19}. Our photometric parameters are consistent to within 2$\sigma$ with the parameters found by \cite{max18} who solved the long-cadence K2 data.  A comparison with physical parameters reported by \cite{rat10} shows remarkably good agreement, although our parameters are an order of magnitude more precise. The masses reported by \cite{syb18} are consistent with our values to within 2$\sigma$, while the radii reported by them are very different. The likely reason of this discrepancy is a use of preliminary light curve of FM Leo from the Solaris project by \cite{syb18} (K. He{\l}miniak -- private communication). We estimate the spectral type of the system to be F6\,V + F6\,V, which compares well with the estimate of F7\,V from \cite{hou99}.

{\it The third light.}
We investigated possible sources of the positive third light in the {\it Kepler} light curve. It would correspond to a star with a brightness of $R\sim13.9$ mag. Within a radius of 12$\arcsec$ around FM~Leo there is no source of light brighter than $R\sim$ 20 mag in the {\it Gaia} DR2 catalogue nor in any other deep catalogue of stellar sources. If this third light comes from a physically bound companion, it would have an absolute magnitude of $M_R=8.0$ mag, corresponding to a K9\,V star. We looked for a sign of the third light in our highest-S/N HARPS spectra taken around quadrature using the broadening function. We used a template corresponding to a K9\,V star. In the red part of the spectrum there is no trace of absorption lines from a possible companion to within a limit of $\sim 0.5$\% of the total light of the system.

FM~Leo was reported to have a quite significant proper motion anomaly \citep{ker19} and was flagged as ``a binary''. In our case it would mean ``a triple'' with a relatively short tertiary orbital period. However, orbital phases calculated separately from radial velocity curves and the light curve are in perfect agreement. Also our analysis of the $O-C$ diagram (Section~\ref{sec:oc}) does not show any trace of a change of the orbital period.  We can also consider one other possibility, that a use of approximate limb darkening laws representing changes of the surface brightness over a stellar disk leads to the residuals. As we noted above, a shift from the logarithmic law to the square root law helped to reduce the systematic residuals in eclipses. The version 2007 of the WD code unfortunately does not provide fitting of the limb darkening coefficients of two-parametric laws, so for the time being we leave open the question of the source of the residuals. For the purpose of the current work we have assumed that the third light is real and comes from a K9\,V companion star on a wide orbit.

\begin{figure}
\hspace*{-0.5cm}
\includegraphics[angle=0,scale=0.55]{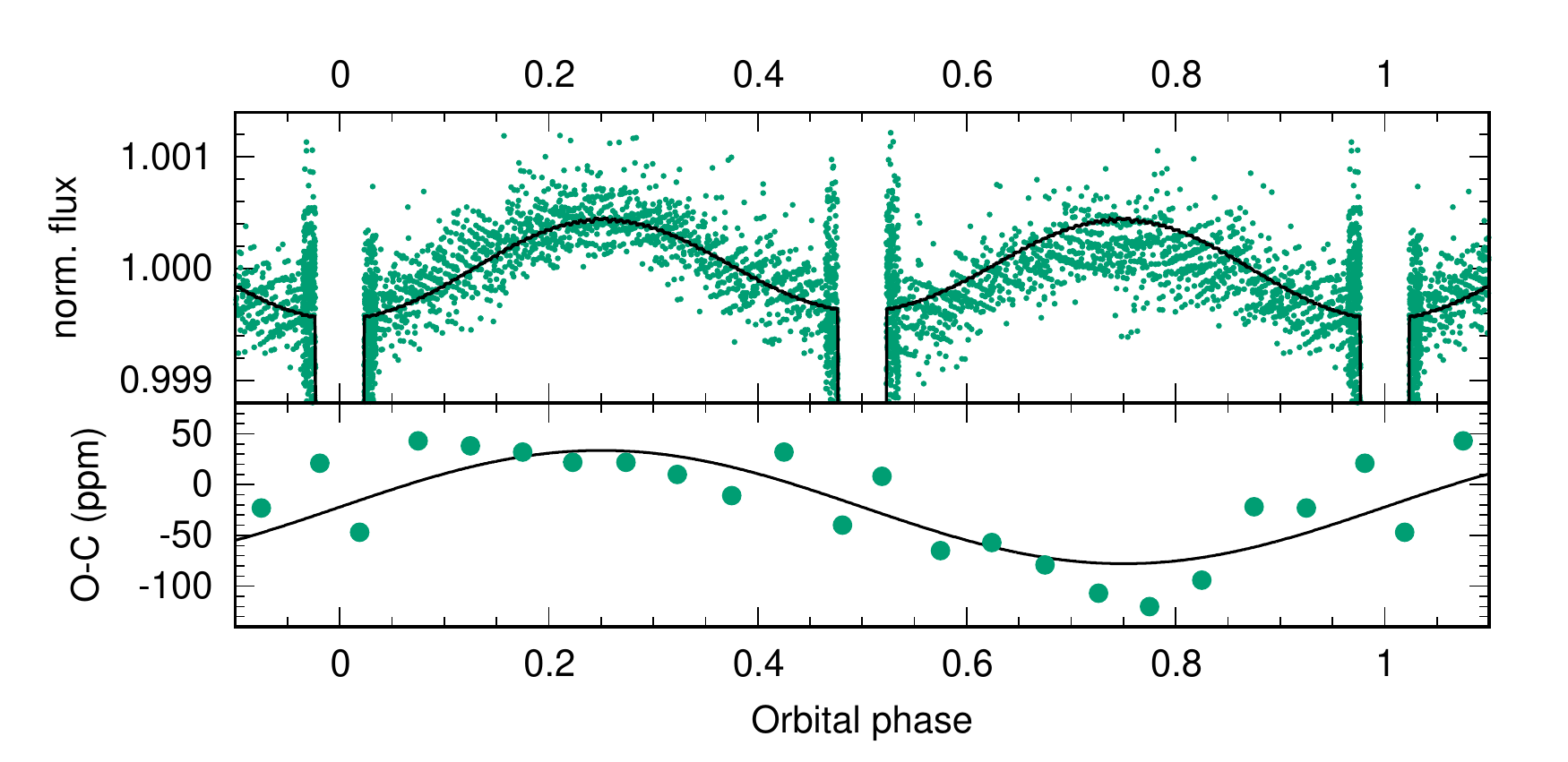}
\caption{{\it Top}: out-of-eclipse flux changes of FM~Leo, with the best fit model from WD overplotted. The ellipsoidal modulation and the reflection is included in the model. {\it Bottom}: residuals binned to steps of 0.05 in phase. The theoretical light variation due to Doppler beaming is overplotted.\label{DopBin}}
\end{figure}

{\it The Doppler beaming.}
The upper panel of Fig.~\ref{DopBin} shows a close-up of the out-of-eclipse flux changes of FM~Leo. Although the best model represents the flux changes very well, it is noticable that during the first quadrature FM~Leo is on average brighter by $\sim$0.1~mmag than during the second quadrature. As the WD model includes flux changes due to tidal deformation of the components and the mutual reflection, persistent systematic residuals of about 120~ppm come most probably from Doppler boosting. To calculate the expected Doppler boosting signal we used equation 9 from \cite{plac19}. We estimated the beaming factor for $T_{\rm eff}=6300$~K and $\log{g}=4.0$ to be $B=3.6\pm0.2$ from fig.~2 in \cite{plac19}. The flux changes due to the Doppler boosting are plotted in the bottom panel of Fig.~\ref{DopBin} together with the binned residuals. The agreement is remarkable, suggesting that the detection is real and significant at the 3$\sigma$ level. To our knowledge this is the first detection of the Doppler beaming in a binary system consisting of two similar main-sequence stars.

{\it Short cadence versus long cadence data.}
Having both sets of data for FM~Leo we made a high precision test of solving independently both sets of photometry and comparing the results. We solved the full long cadence data for FM~Leo in a similar manner to the short cadence data. In order to account for the phase smearing which flattens out the brightness variations we used an iterative approach to correct the long cadence light curve (LCLC). In case of FM~Leo one photometric point of LCLC corresponds to the integration time of $\sim$0.003 orbital phases. That is a significant part of the eclipse duration: almost 7\%. Based on the preliminary WD model obtained from LCLC, we determined corrections as follows. We generated a synthetic normalised light curve with very dense coverage of the orbital phase (5000 points). Based on this for every synthetic point we calculated a mean synthetic flux corresponding to the phase smearing of 0.003. The difference of the two is a correction which we applied to every point of the phased LCLC. Then we solved the corrected light curve to derive an improved model. We iterated this scheme three times until the resulting differences between the new and old corrections were below 200~ppm.

Both model solutions, one obtained by fitting directly the short cadence light curve (SCLC) and one obtained by fitting the corrected LCLC are in perfect agreement: not only are the synthetic light curves essentially the same but also the resulting photometric model parameters are the same to within the uncertainties. As already pointed out by \cite{max18}, we note that the residuals of the LCLC solution during eclipse are twice as high as the out-of-eclipse residuals. However, the SCLC light curve shows the same magnitude of residuals within and outside eclipse. We conclude that this iterative scheme can be reliably applied to solve K2 and TESS light curves of eclipsing binaries having only long cadence data.

\subsubsection{BN~Scl}
\label{full:bnscl}
\begin{figure}
\hspace*{-0.5cm}
\includegraphics[angle=0,scale=0.55]{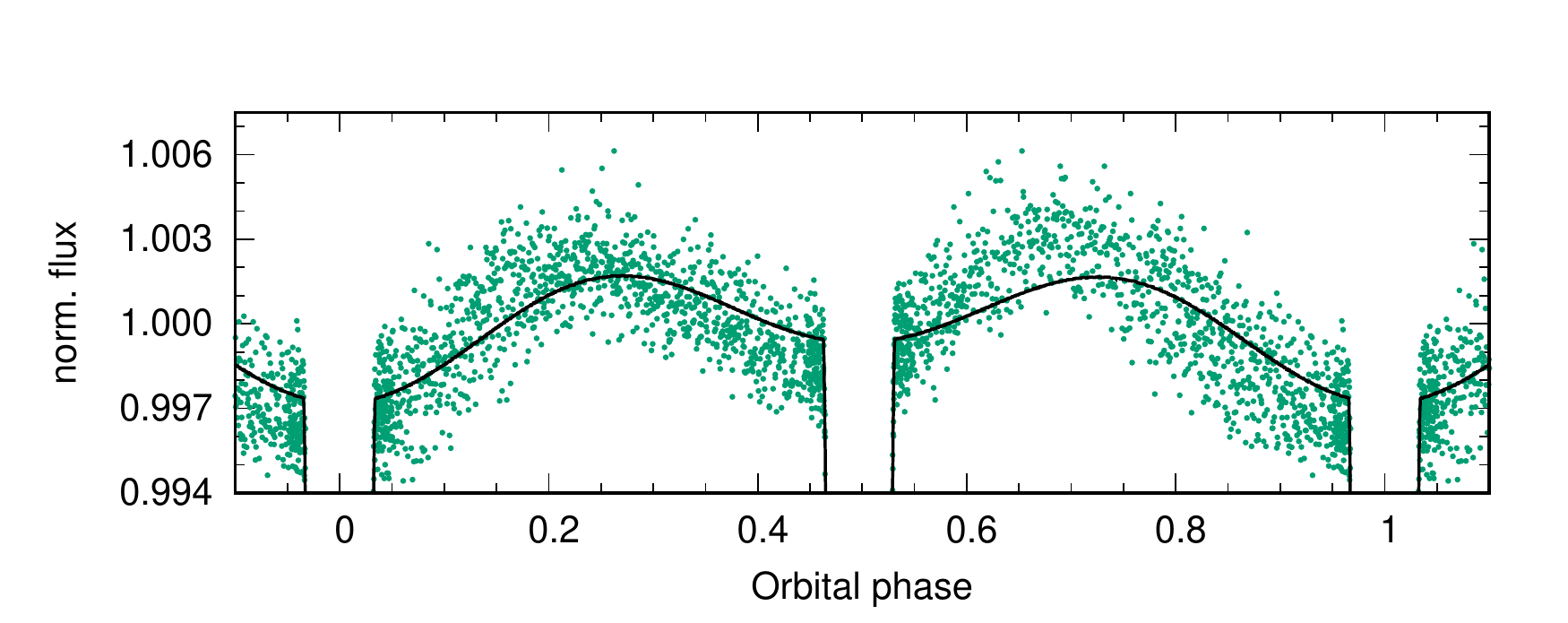}
\caption{Out-of-eclipse flux changes of BN~Scl, with the best fit model from WD overplotted. The deviations from the best fit at the orbital phases from 0.4 to 0.8 due to the spot activity can reach $\sim$1.5 mmag. \label{bn:out}}
\end{figure}

Because of the small spots affecting the light curve of BN~Scl, we used only a part of TESS photometry when the spot activity was the smallest. We did not fit spots on any of the components, instead we allowed to fit both albedo parameters $A_1$ and $A_2$. We used the grid fineness parameters \verb"N1=N2"=60 and the logarithmic limb darkening law. The photometry and velocimetry were solved simultaneously. The albedo of the primary is very low and consistent with zero, see Table~\ref{tab_par_orb}, most probably because of the presence of the spots. The best fit model presented in Fig.~\ref{fig:light} still shows the small systematic residuals of $\sim$ 500 ppm near mid-eclipse during the secondary minimum. A use of the square root law for the limb darkening does not provide any improvement to the solution. There is a faint, red star $\sim$8 arcsec to the south-west of BN~Scl which is on the edge of the TESS photometric aperture. This is TIC 155543721, with an estimated TESS magnitude of 15.8 (0.1\% of BN~Scl's flux).  Allowing the third light to be a free parameter of the light curve fit does not help to reduce the residuals and the returned values of $l_3$ are consistent with 0. We put the upper limit for the $l_3$ as 0.1\% in the TESS light curve. Most probably the presence of small spot(s) affects the model fit and produces small systematic residuals at the secondary minimum (see Fig.~\ref{bn:out}).

The components are slightly evolved, especially the primary. The tidal distortion amounts to 0.6\% for the primary and 0.3\% for the secondary.  The system may be very valuable for testing models of stellar evolution as the components differ significantly in their masses and are below the 2~M$_\odot$ limit where core overshooting may depend strongly on  mass. The metallicity of the system is significantly sub-solar. The spectral types of components are F7\,V + F8\,V, in a perfect agreement with the estimate of F7\,V from \cite{hou82}.

\section{Calibration of new surface brightness - colour relations.}\label{sbcr}

In our previous work \citep{gra17} we investigated the surface brightness -- colour relations derived solely from eclipsing binary stars using a mixture of {\it Hipparcos} \citep{vLe07a} and {\it Gaia} DR1 parallaxes \citep{gaia16}. Here we derive new relations utilising much more precise {\it Gaia} EDR3 parallaxes \citep{gaia20}.

\subsection{Sample}

We used a compilation of eclipsing binary stars from \cite{gra19}, with updated parameters on three systems: AL~Dor, AL~Ari and FM~Leo (this paper). To the sample we added four new systems: BN~Scl, $\beta$~Aur, NN~Del and V1022~Cas. In the case of $\beta$ Aur we used parameters from \cite{sou07} and \cite{beh11} while for NN Del we adopted masses from \cite{gal19} and photometric parameters from \cite{syb18}. For AI~Phe we adopted very precise parameters from recent papers by \cite{gal19} and \cite{max20}. Parameters for V1022 Cas were adopted from a number of recent works \citep{fek10,les19,sou20}. In order to minimise the interstellar extinction and the {\it Gaia} zero-point uncertainties we set a distance limit of 200 pc corresponding to the trigonometric parallaxes $\varpi>5$ mas. We also put a precision limit of 1.5\% on the relative radii of both components, resulting in a sample of 27 systems. For each we downloaded the {\it Gaia} EDR3 parallaxes and auxiliary parameters from the {\it Gaia} Archive\footnote{\texttt{https://gea.esac.esa.int/archive/}}. None of selected systems has a \verb"RUWE" parameter\footnote{\texttt{https://gea.esac.esa.int/archive/documentation/\\GDR2/Gaia\_archive/chap\_datamodel/sec\_dm\_main\_tables/\\ssec\_dm\_ruwe.html}} larger than 1.4, which would signify a problematic astrometric solution. For $\beta$~Aur we used the \textit{Hipparcos} parallax \citep{vLe07a}. In the final sample of 28 systems the median errors on the parallaxes and absolute radii are 0.3\% and 0.6\%, respectively.

\subsection{Geometric distances}
{\it Gaia} EDR3 significantly improves over {\it Gaia} DR2 in terms of the precision and the accuracy of the astrometry \citep{lind20b}. In particular, the zero-point shift of {\it Gaia} EDR3 parallaxes (the parallax bias) was determined \citep{lind20a} as a function of the $G$ magnitude (\verb"phot_g_mean_mag" in the {\it Gaia} Archive), a colour and an ecliptic latitude $\beta$ (\verb"ecl_lat"). As a colour parameter, rather than using the colour index $G_{\rm BP} - G_{\rm RP}$, an effective wavenumber $\nu_{\rm eff}$ (\verb"nu_eff_used_in_astrometry") was used in the five-parameter astrometric solutions and an astrometric pseudo-colour $\hat{\nu}_{\rm eff}$ (\verb"pseudocolour") in the six-parameter astrometric solutions. In {\it Gaia} EDR3 quasars have a median parallax of $-17$ $\mu$as and in general brighter and bluer objects have a larger negative parallax bias. The bias function $Z(G,\nu_{\rm eff},\beta)$ is provided separately for the five- ($Z_5$) and six-parameter ($Z_6$) solutions as a Python module \verb"zero-point", available via the {\it Gaia} web pages\footnote{\texttt{https://www.cosmos.esa.int/web/gaia/EDR3-code}}.

In order to calculate the new surface-brightness colour relations we used parallaxes corrected for the parallax bias. The average change of parallax in our sample due to this correction is just $-$0.3\%, thanks to the proximity of the systems ($\varpi>5$ mas).

\subsection{Photometry}
We used our compilation \citep{gra19} as a source of optical Johnson $B$ and $V$ magnitudes. For new systems we used mean magnitudes from a number of catalogues available in the VizieR catalogues access tool. $G_{BP}$ and $G$ EDR3 photometry was downloaded from the {\it Gaia} Archive. For near-infrared photometry we used 2MASS magnitudes from \cite{cut03} which were corrected for the presence of eclipses. The corrections were calculated from synthetic light curves in $JHK$ bands using our WD models of the systems. In case of $\beta$~Aur we used Johnson photometry from a compilation by \cite{duc02} transformed into the 2MASS photometric system.

\begin{figure*}
\centering
\includegraphics[width=.34\textwidth]{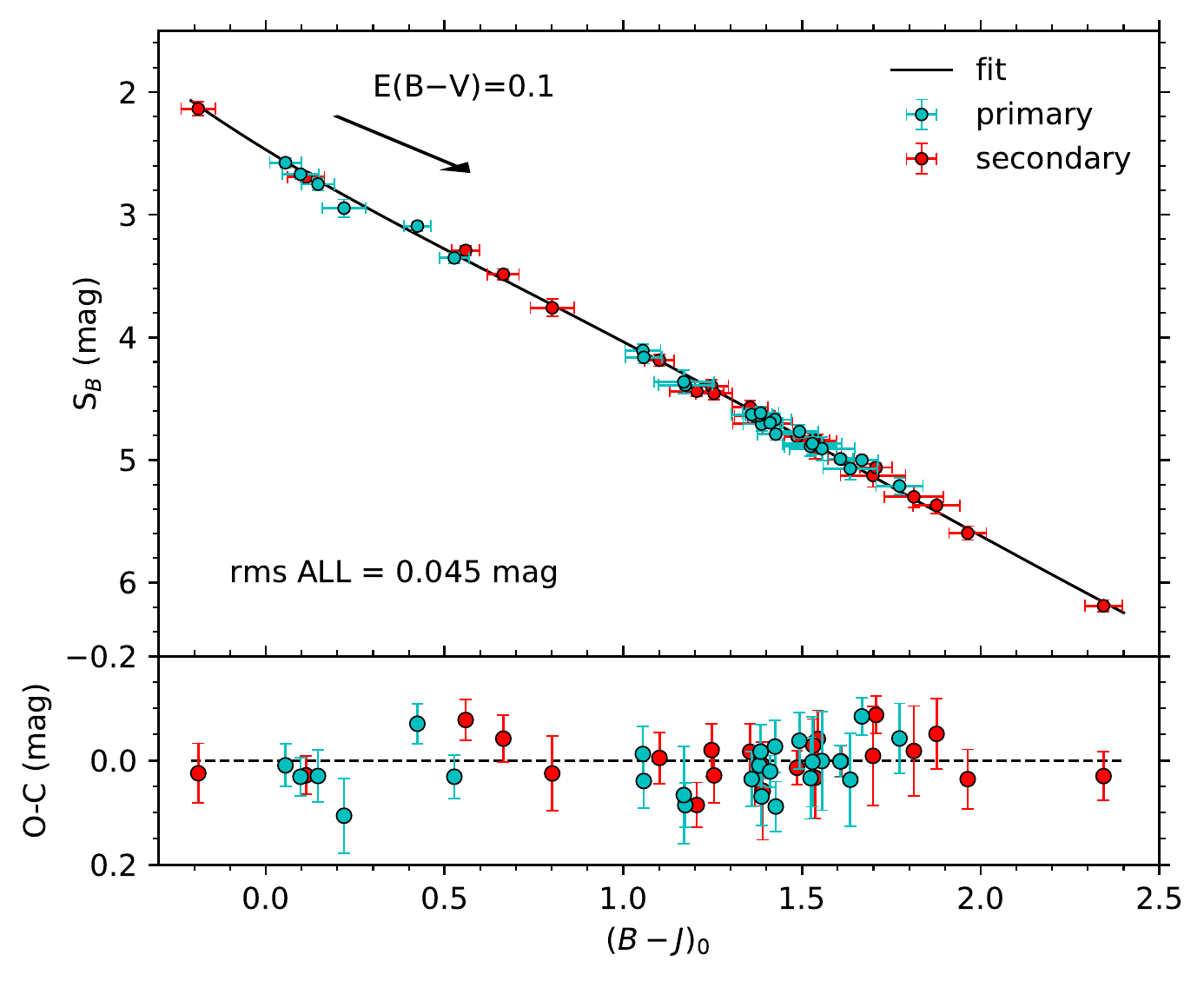}\hspace*{-0.19cm} %\hfill%
\includegraphics[width=.34\textwidth]{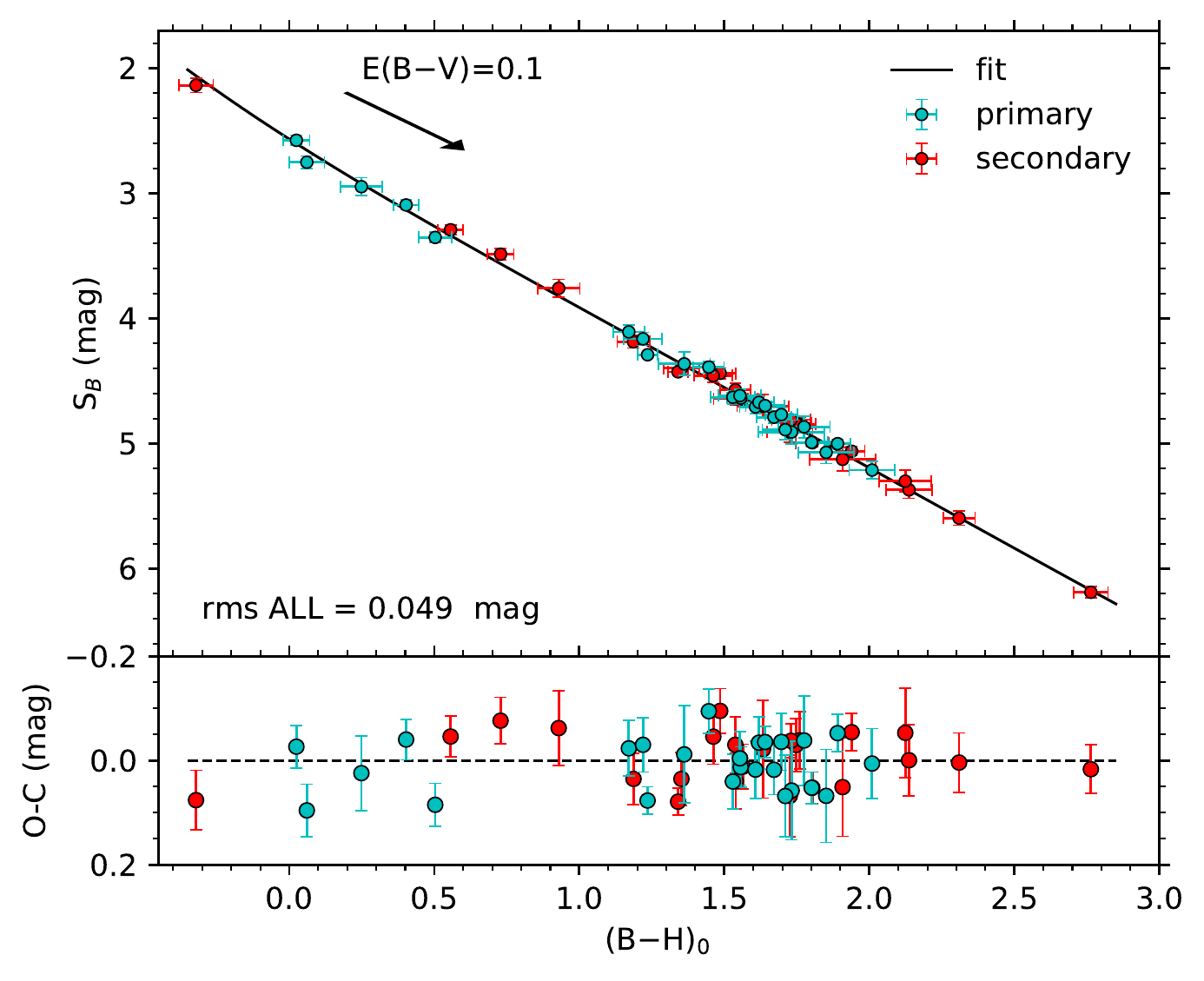}\hspace*{-0.12cm}%\hfill%
\includegraphics[width=.34\textwidth]{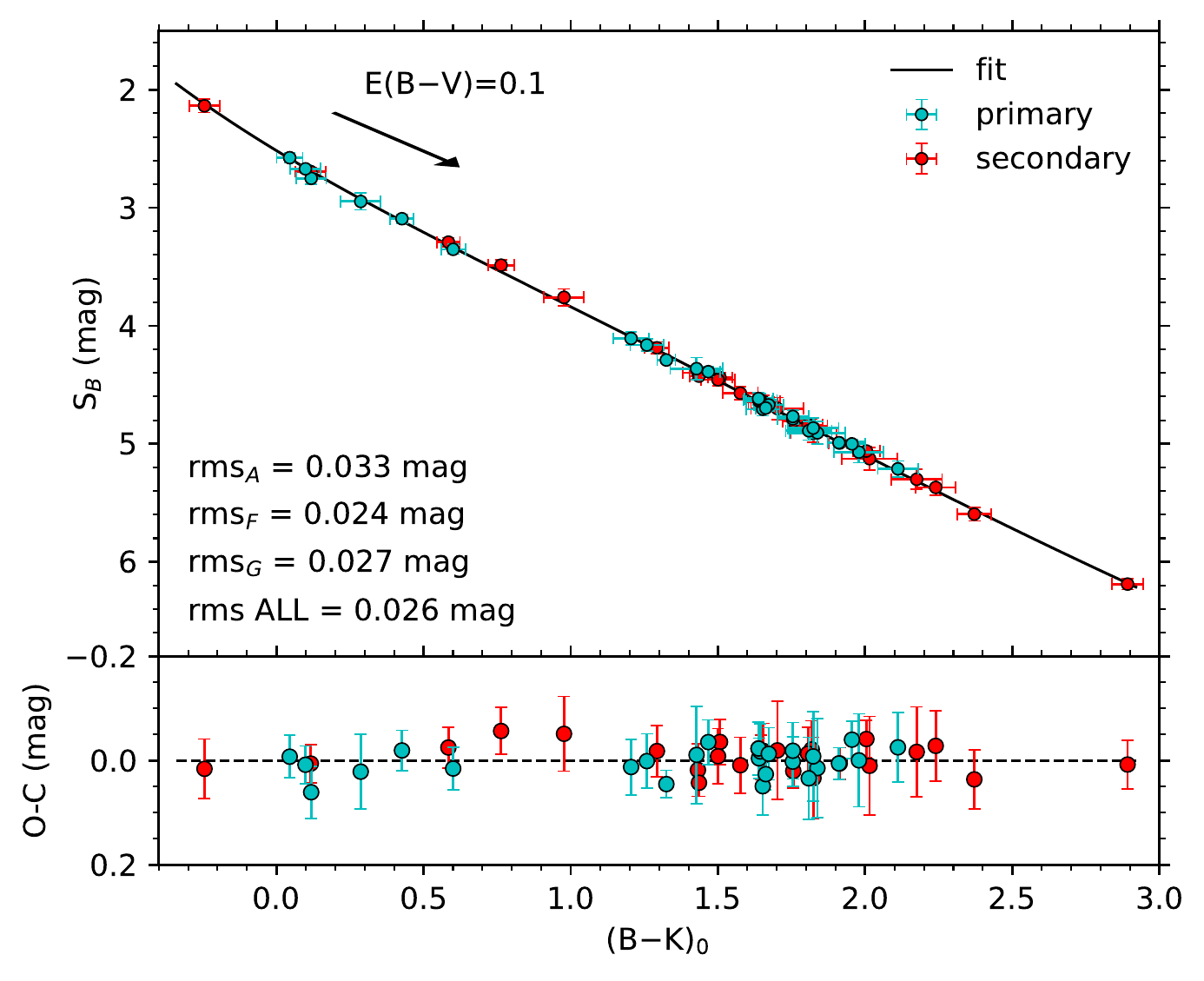}\\
\includegraphics[width=.34\textwidth]{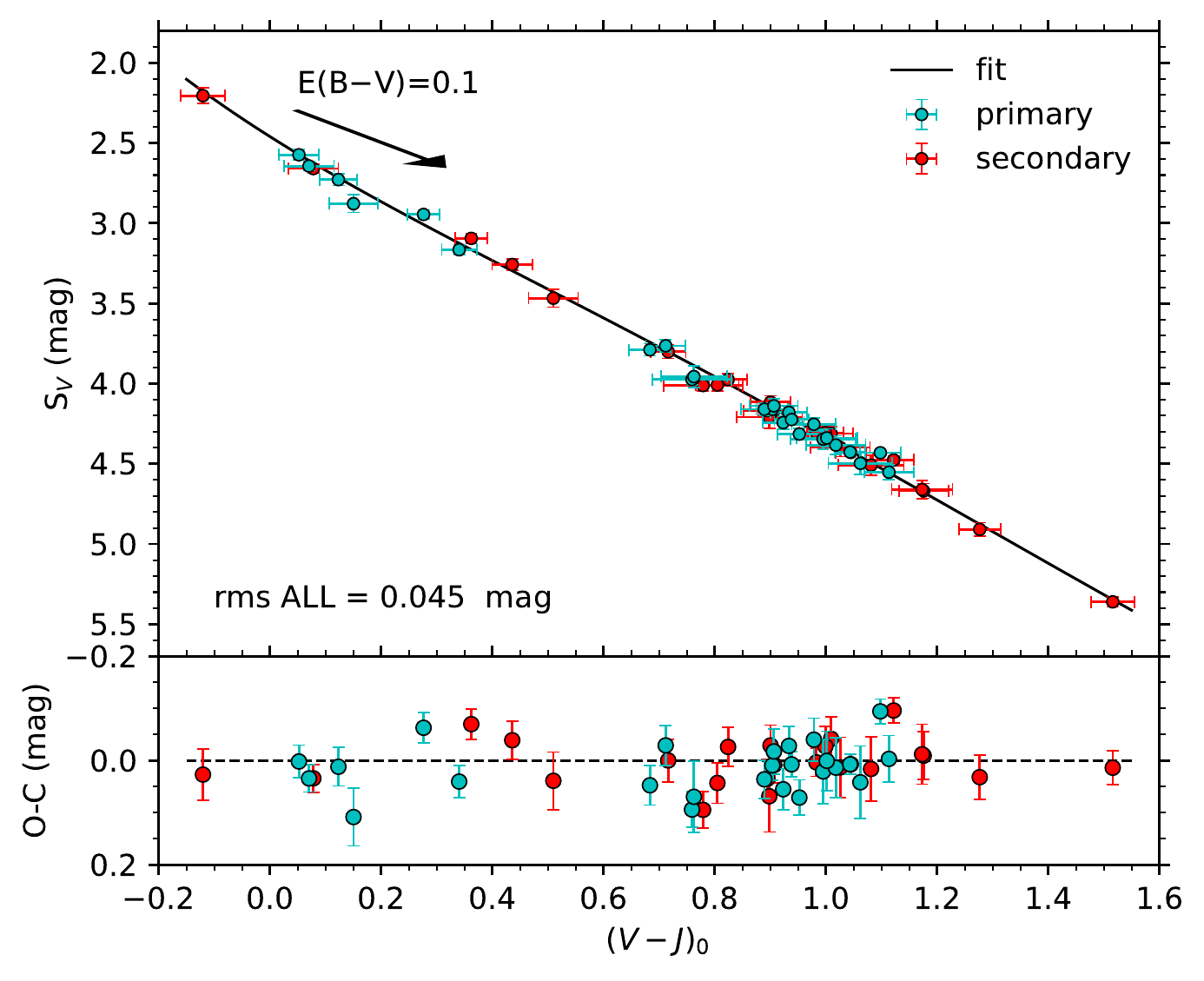}\hspace*{-0.19cm} %\hfill%
\includegraphics[width=.34\textwidth]{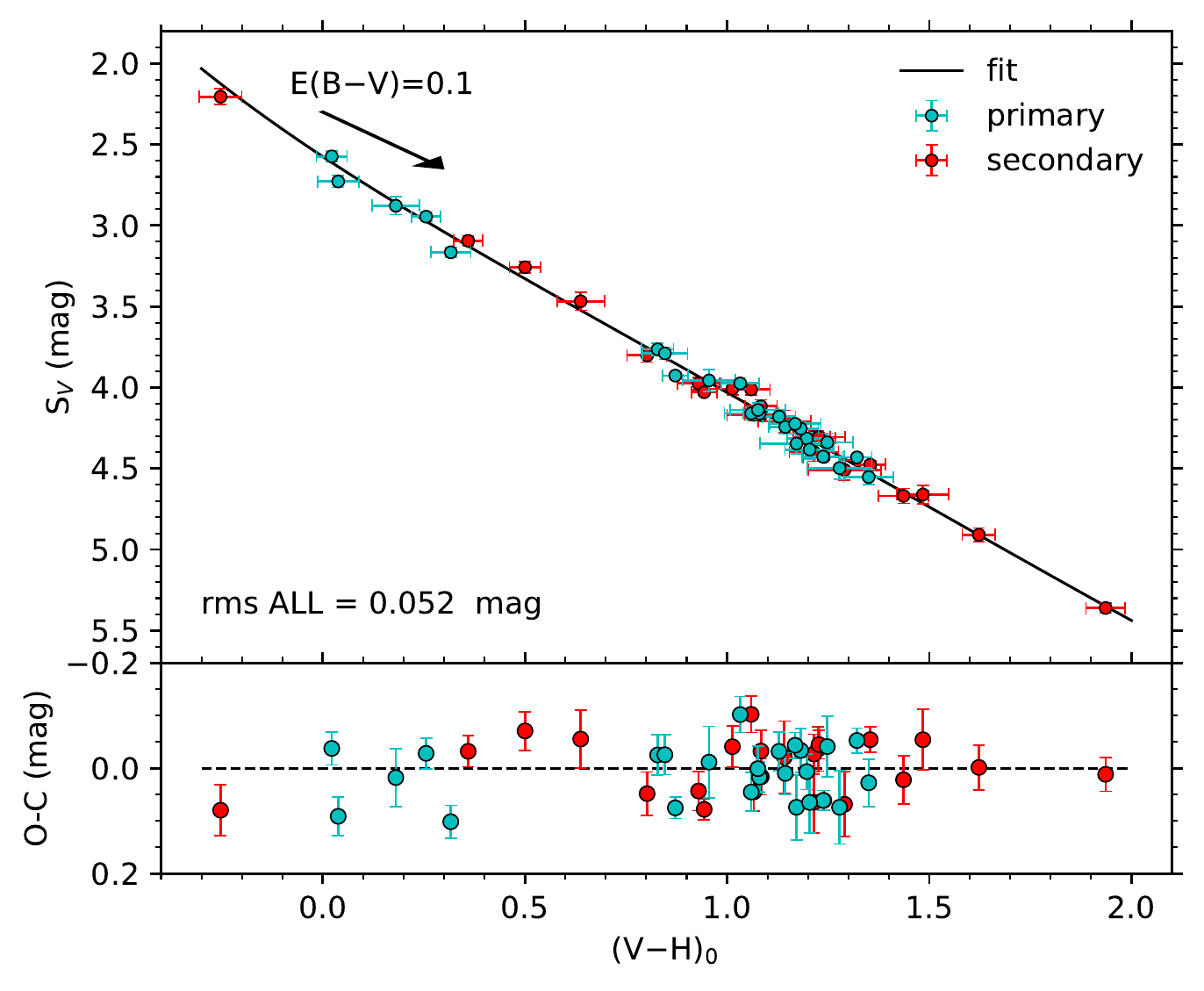}\hspace*{-0.12cm}%\hfill%
\includegraphics[width=.34\textwidth]{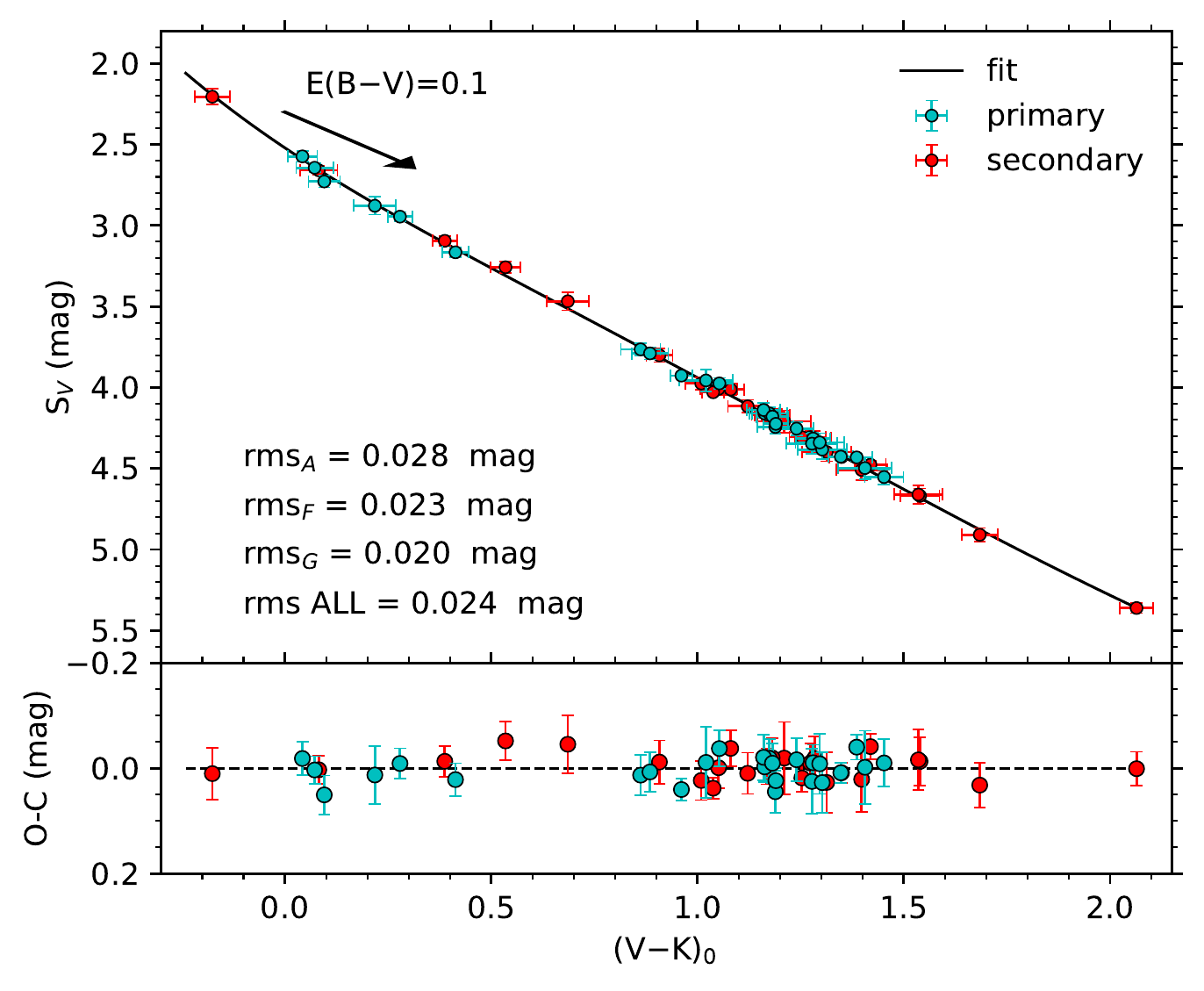}\\
\includegraphics[width=.34\textwidth]{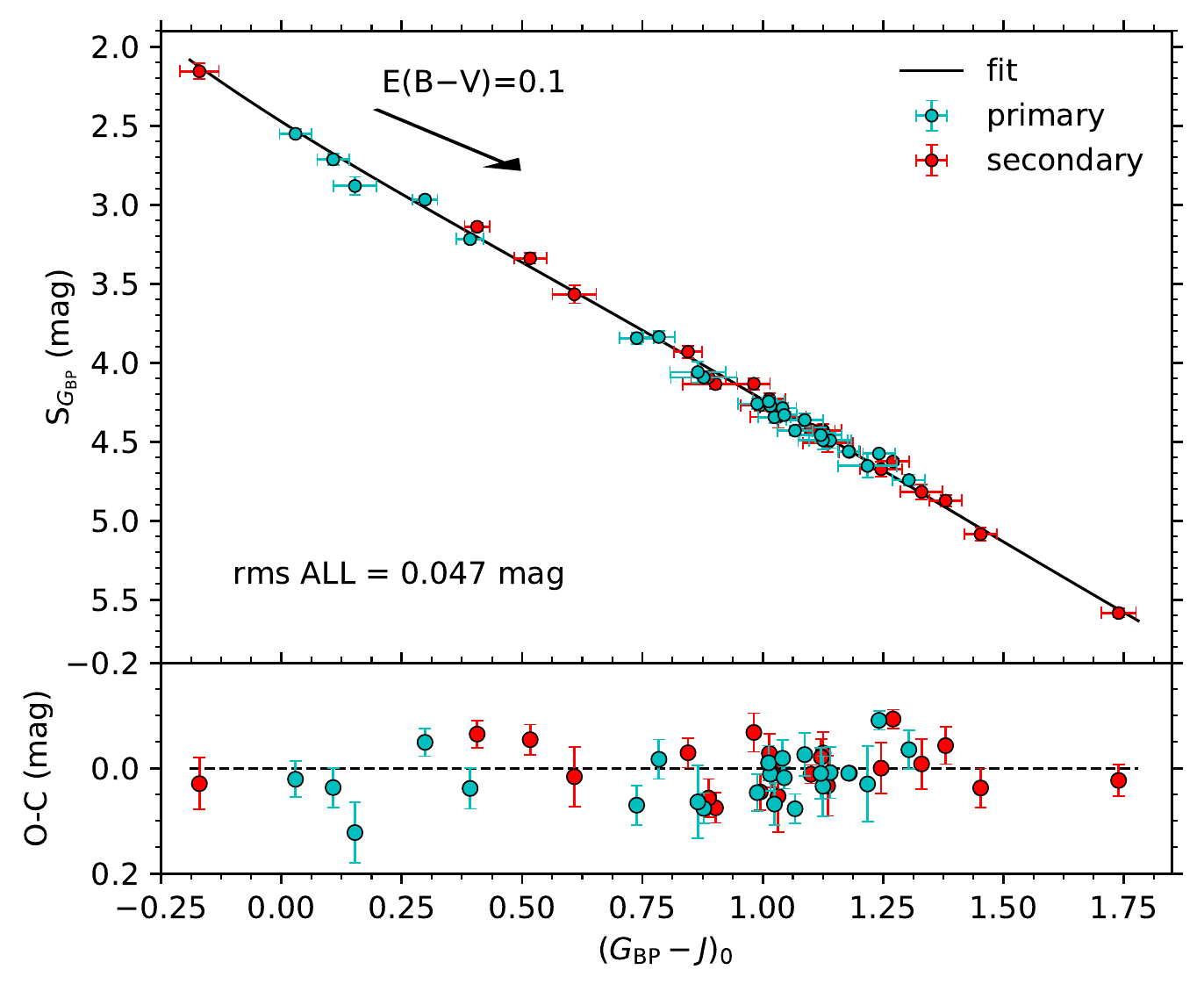}\hspace*{-0.19cm} %\hfill%
\includegraphics[width=.34\textwidth]{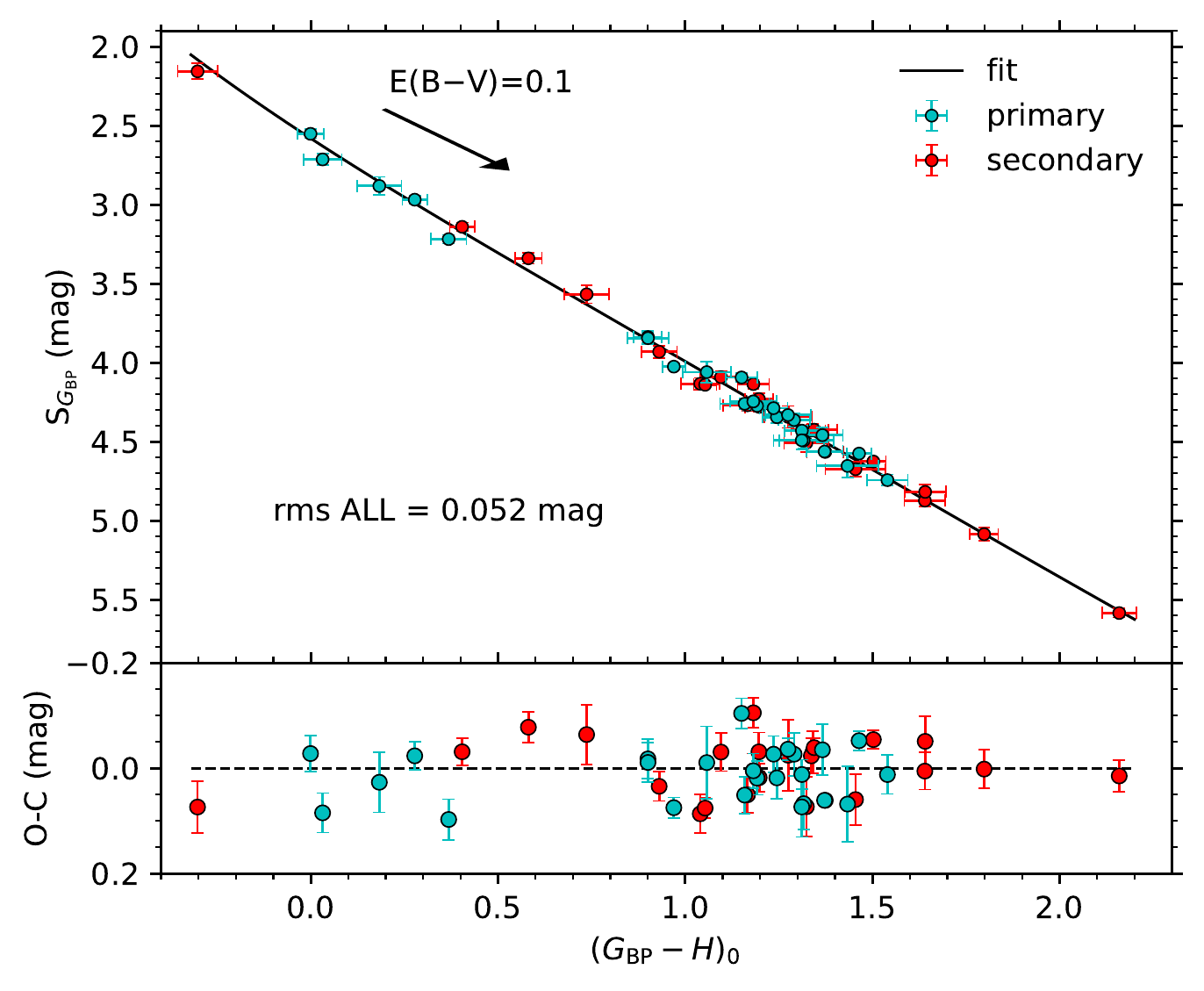}\hspace*{-0.12cm}%\hfill%
\includegraphics[width=.34\textwidth]{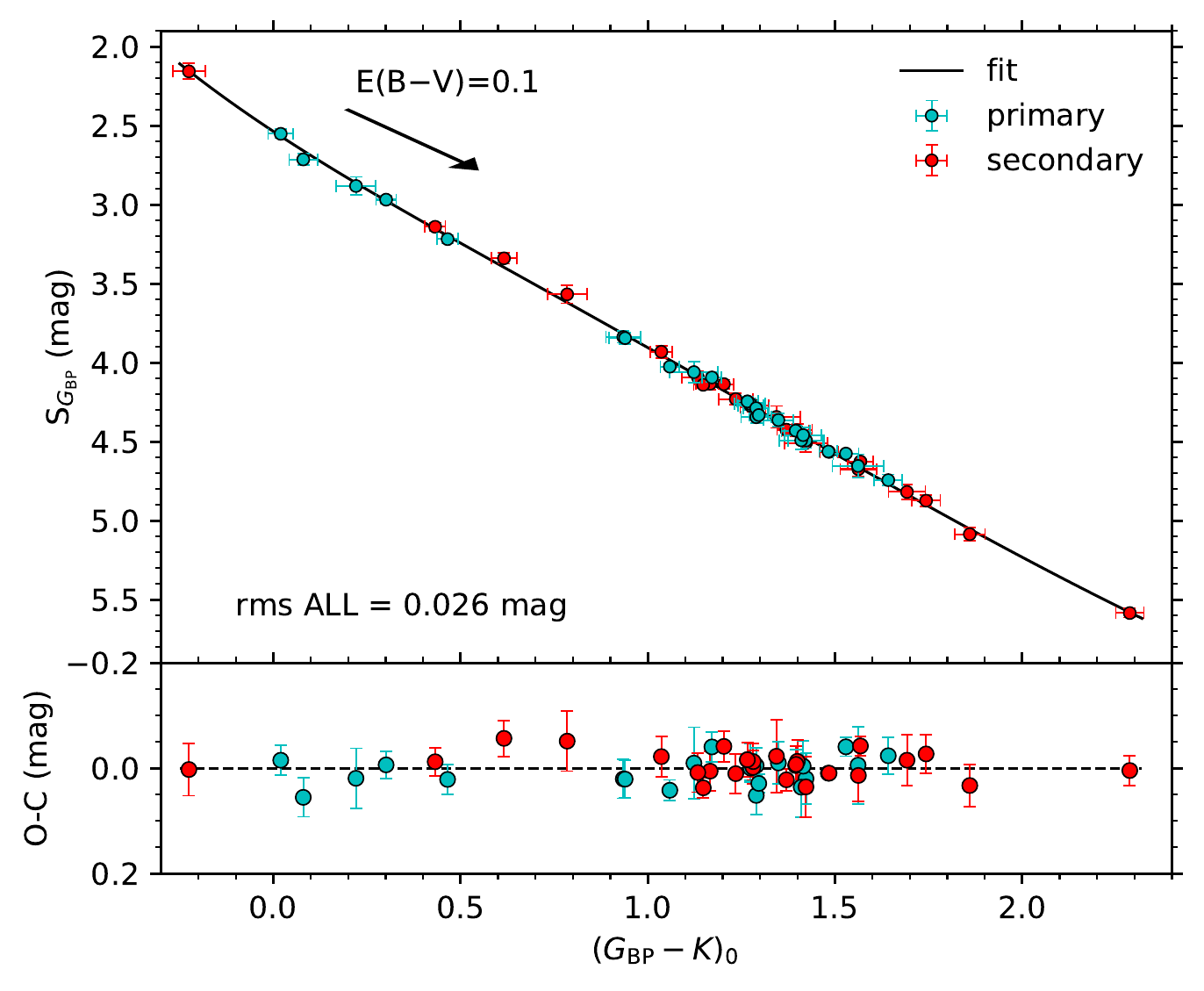}\\
\includegraphics[width=.34\textwidth]{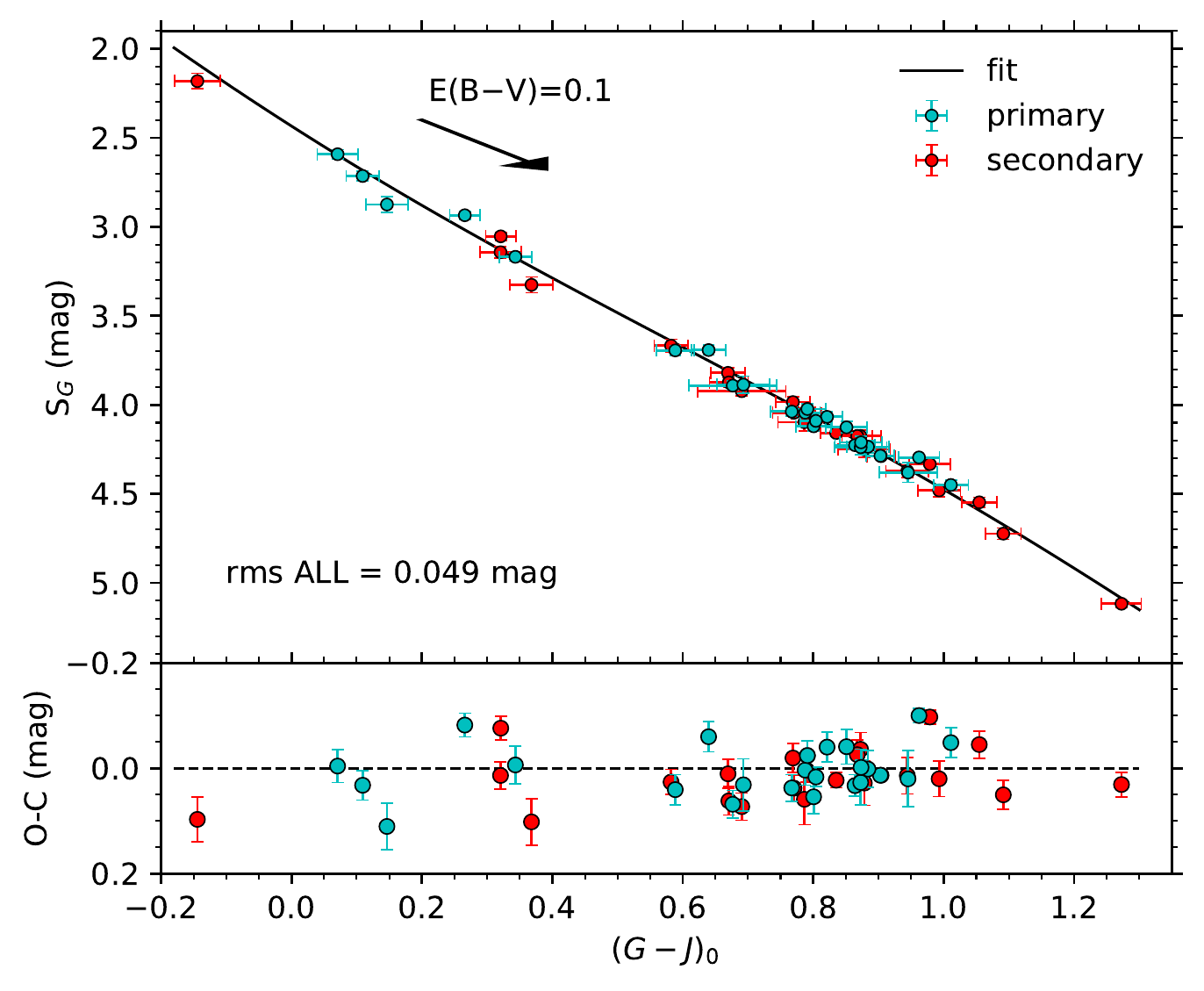}\hspace*{-0.19cm} %\hfill%
\includegraphics[width=.34\textwidth]{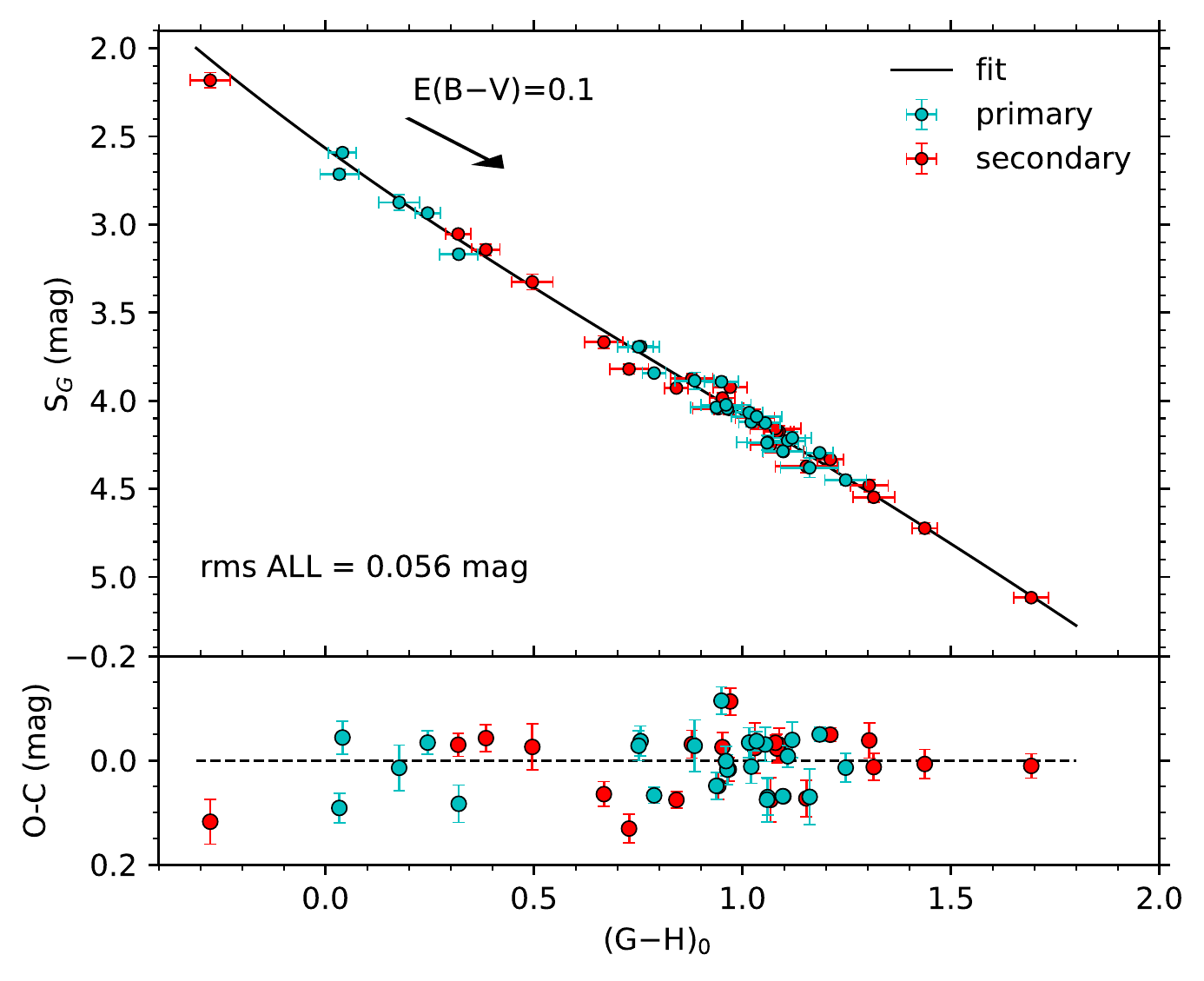}\hspace*{-0.12cm}%\hfill%
\includegraphics[width=.34\textwidth]{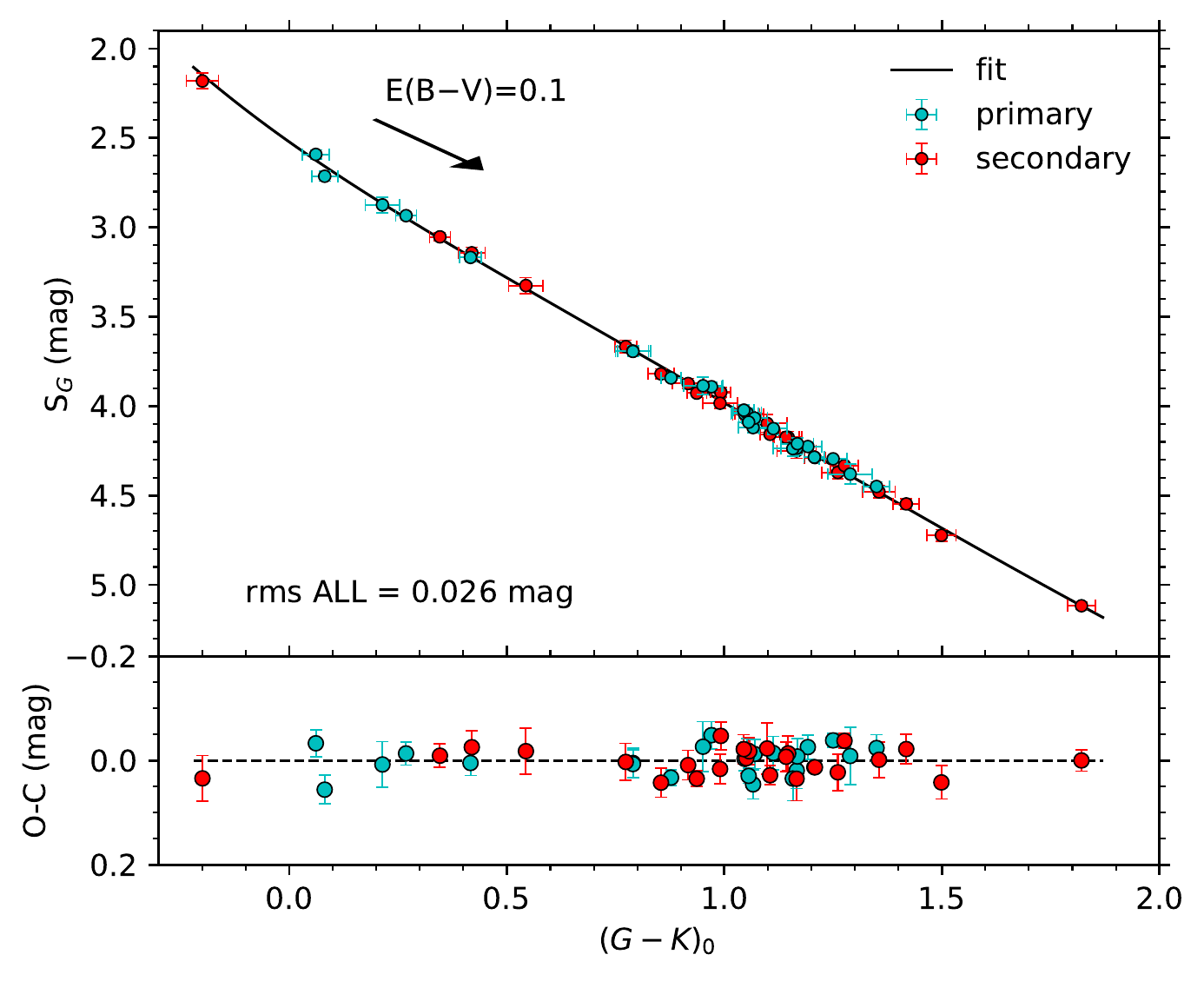}
\caption{The SBCRs derived for eclipsing binary stars and calibrated for Johnson $B$, $V$ filters and {\it Gaia} EDR3 $G_{\rm BP}$, $G$ filters. Infrared magnitudes are expressed in the 2MASS system. \label{fig:sbcr}}
\end{figure*}

\subsection{The third light}
Intrinsic colours of the components were calculated in each band from the light ratios we determined using the WD code. However, in few cases we had to account for the third light. Two systems are confirmed triples: AI~Phe and AL~Ari. A characteristics of the third light in the TESS light curve of AI~Phe was presented by \cite{max20}, with a suggestion that a companion is a late-K dwarf. For this paper we assumed that the companion is a K9\,V star and we subsequently computed its contribution to the various bands using an online table of intrinsic colours of main-sequence stars\footnote{\texttt{http://www.pas.rochester.edu/$\sim$emamajek/\\EEM\_dwarf\_UBVIJHK\_colors\_Teff.txt}} \citep{pec13}. The expected contribution of the companion to the total flux in the $K$ band is 1.6\%, and in the optical is less than 0.2\% ($B$ band).

In the case of AL~Ari we can only put an upper limit on the third light of 1\% in the $V$ band. At the distance of AL~Ari ($\sim$140~pc) this means that the companion's spectral type would be later than K8\,V. However, the proper motion anomaly \citep{ker19} and the small acceleration of the system's barycentre suggest a lower-mass companion. For the purpose of this work we do not assume any parameters for the companion and we set $l_3=0$ in all bands.

Another system, FM~Leo, is a suspected triple because of the third light, which we detected in the K2 light curve and by its significant proper motion anomaly. We calculated the contribution of the putative companion star (a K9 dwarf) in different bands in a similar manner to AI~Phe. In the $K$ band the expected contribution amounts to $3.3\%$ of the total flux of the system, while in the $B$ band only to 0.15\%. For both AI~Phe and FM~Leo we used these contributions to correct the intrinsic colours and the surface brightness parameters of their components.

\subsection{Results}

Our sample of eclipsing binaries covers stars with spectral types from B7\,V (the primary of GG~Lup) down to M1\,V (the secondary of V530~Ori). However, these extremes are much bluer and much redder than the rest of the sample and all calibrations were done in smaller colour ranges corresponding to spectral types ranging from B9\,V (the secondary of GG\,Lup) down to K0\,IV (the secondary of AI Phe). All components are main-sequence stars with the exception of very few subgiants. In the sample the mean surface gravity is $\log{g}=4.22$ and the lowest value is $\log{g}=3.6$.

The surface brightness parameter $S$ was calculated for each band using equation 5 from \cite{hin89}:
\begin{equation}
\label{equ:sb}
S_{\!m} = 5 \log{\theta_{\rm LD}} + m_0,
\end{equation}
where $m_0$ is the intrinsic magnitude of a star in given band and $\theta_{\rm LD}$ is the limb darkened angular diameter expressed in milliarcseconds. The angular diameters were calculated using:
\begin{equation}
\theta_{\rm LD} =  9.301\cdot 10^{-3}\, R\, \varpi_{Gaia},
\end{equation}
where $R$ is the stellar radius expressed in nominal solar radii $\mathcal{R}_\odot$ \citep{prsa16}.

The colours of the components were calculated from the dereddened observed magnitudes and the light ratios in different bands provided by the WD models. Because for a spectral type $\sim$A0 the SBCRs show significant nonlinearity \citep[see e.g.][]{cha14}, we decided to fit fifth-order polynomials. Fits were done using the orthogonal distance regression. Fig.~\ref{fig:sbcr} shows calibrations of the surface brightness parameter $S$ in four filters ($B$, $V$, $G_{\rm BP}$ and $G$) against the optical-infrared colours. All SBCRs calibrated against the 2MASS $K$ band show the smallest residuals with {\it rms} of $\sim$0.025 mag ($\sim$1\%). Notably the SBCRs calibrated with the 2MASS $J$ and $H$ bands show considerably larger residuals by a factor of $\sim$2. The likely reason for this is the precision of the 2MASS photometry itself, which reaches 1\% in the $K$ band and only 2\% in the $J$ and $H$ bands. At this moment the quality of the infrared photometry is the main limiting factor to the precision with which we can calibrate the SBCRs. Another significant source of residuals are the radii of eclipsing binary components, which are known with a precision of 1--2\% for a number of systems.

In Table~\ref{tab:calib} we present coefficients of the fifth-order polynomial fits over the colour range in which a particular SBCR can be safely used. Instead of giving uncertainties of fitted coefficients we give information about the precision of the predicted stellar angular diameter.

{\it Gaia} EDR3 does not account for binary solutions so systems with larger photocentre movement (PHM) may have their parallaxes affected, and this will contribute to the scatter in the SBCRs. In our sample ten systems have significant amplitudes of the PHM, i.e.\ they are much larger than the errors of parallaxes. \cite{gal19} determined precise orbital parallaxes for four systems in our sample using interferometry, and these are consistent to within 1$\sigma$ with {\it Gaia} parallaxes for AL~Dor and AI~Phe (which have an insignificant PHM) but disagree at the 2.5$\sigma$ level for KW~Hya and NN~Del (which have a significant PHM). On the other hand, only NN~Del has a large PHM amplitude (8\% of the {\it Gaia} parallax), whilst for the rest of the sample the amplitude is smaller than 3\% with a median of just 0.6\%.

\begin{table*}
\centering
\caption{Coefficients of polynomial fits to the surface brightness parameter $S$ in Johnson $B$, $V$ and {\it Gaia} $G_{\rm BP}$, $G$ bands for dwarfs and subgiants.  \label{tab:calib}}
\begin{tabular}{@{}lcccccccccc@{}}
\hline \hline
Band & Colour & Number & Valid over & $a_0$ &$a_1$ & $a_2$&$a_3$ &$a_4$ & $a_5$ & $\sigma$ \\
 & $X$  & of Stars & Colour Range & && & & &  & $\%$\\
 \hline
\multicolumn{11}{c}{Fifth order polynomial fits}\\
$B$&$(B\!-\!J)$&52&[$-$0.2:2.4]&2.471&1.786&$-$0.499&0.386&$-$0.121&0.0131&2.1\\
$B$&$(B\!-\!H)$&52&[$-$0.35:2.8]&2.566&1.483&$-$0.236&0.127&$-$0.0313&0.0280&2.3\\
$B$&$(B\!-\!K)$&54&[$-$0.3:2.9]&2.516&1.515&$-$0.358&0.220&$-$0.0597&0.0056&1.2\\
$V$&$(V\!-\!J)$&52&[$-$0.2:1.55]&2.460&2.195&$-$1.041&1.143&$-$0.493&0.0720&2.1\\
$V$&$(V\!-\!H)$&52&[$-$0.3:2.0]&2.575&1.649&$-$0.422&0.331&$-$0.117&0.0148&2.4\\
$V$&$(V\!-\!K)$&54&[$-$0.2:2.1]&2.521&1.708&$-$0.705&0.623&$-$0.239&0.0313&1.1\\
$G_{\rm BP}$&$(G_{\rm BP}\!-\!J)$&50&[$-$0.2:1.75]&2.474&1.934&$-$0.524&0.539&$-$0.220&0.0305&2.2\\
$G_{\rm BP}$&$(G_{\rm BP}\!-\!H)$&52&[$-$0.3:2.2]&2.580&1.544&$-$0.272&0.199&$-$0.0673&0.0082&2.4\\
$G_{\rm BP}$&$(G_{\rm BP}\!-\!K)$&52&[$-$0.3:2.3]&2.535&1.566&$-$0.472&0.404&$-$0.148&0.0184&1.2\\
$G$&$(G\!-\!J)$&50&[$-$0.15:1.3]&2.434&2.339&$-$0.579&$-$0.036&0.392&$-$0.146&2.3\\
$G$&$(G\!-\!H)$&52&[$-$0.3:1.7]&2.567&1.717&$-$0.334&0.124&0.0129&$-$0.0086&2.5\\
$G$&$(G\!-\!K)$&52&[$-$0.2:1.9]&2.522&1.724&$-$0.648&0.597&$-$0.249&0.0372&1.2\\
\hline
\end{tabular}
\tablefoot{The $S$ parameter is defined by Equation~\ref{equ:sb}. Infrared magnitudes are in the 2MASS photometric system. Limb-darkened stellar angular diameter $\theta_{\rm LD}$, expressed in milliarcseconds, follows from the equation $\log{\theta_{\rm LD}} =  0.2\times(a_0 - m +a_1\times X+ ... + a_5\times X^5)$, where $m$ is the observed magnitude (extinction-free) of a star in a particular band and $X$ is a colour (extinction-free). The last column gives the precision in predicting the angular diameter of stars in a valid colour range.}
\end{table*}

\subsection{Comparison with previous calibrations}
\begin{figure}
%\hspace*{-0.5cm}
\includegraphics[angle=0,scale=0.55]{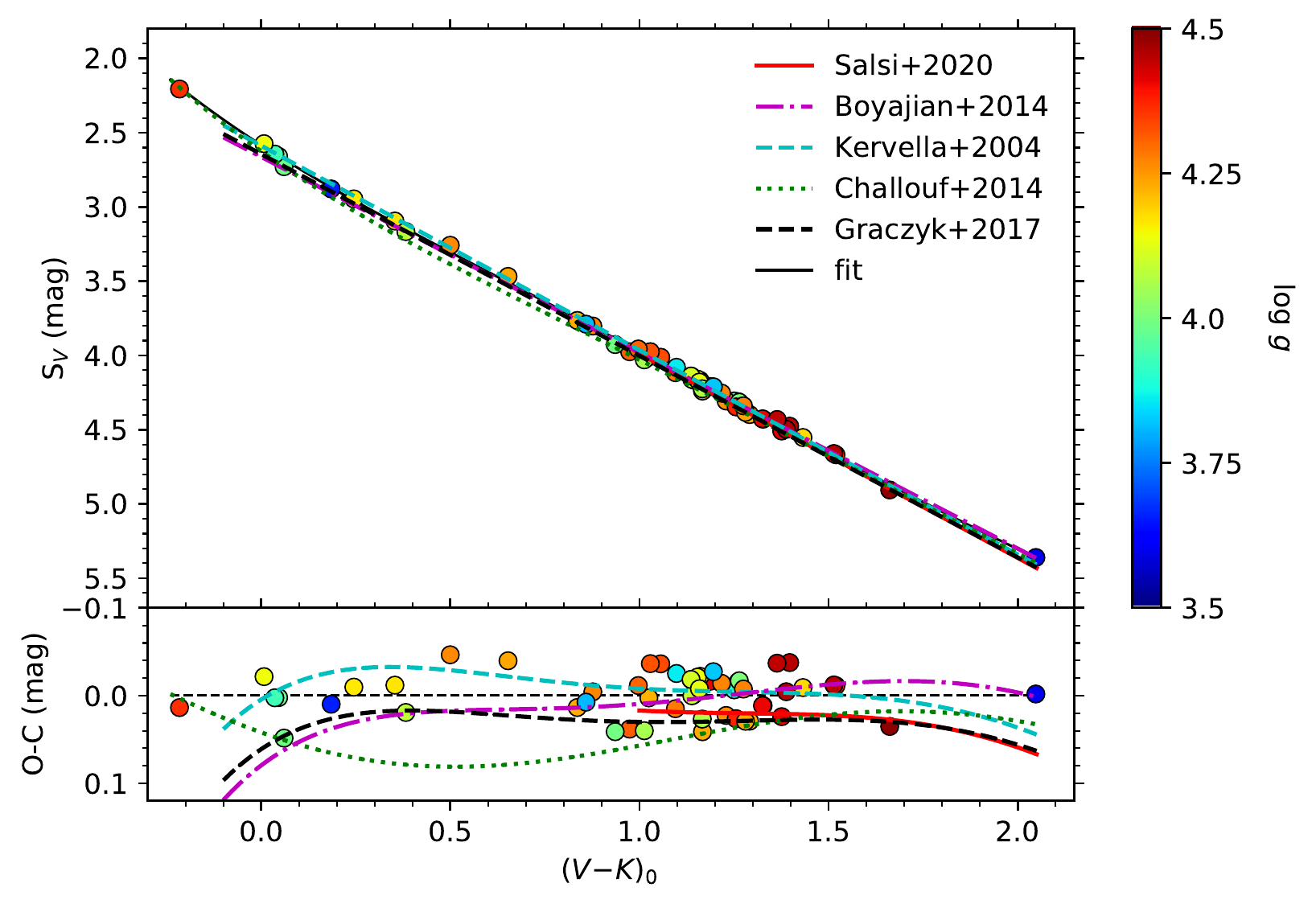}
\caption{Comparison of different $V$-band surface brightness calibrations onto Johnson $(V\!-\!K)$ colour. }
\label{sbcr:comp}
\end{figure}

We compared the SBCR from Table~\ref{tab:calib} for $S_V$ and $(V\!-\!K)$ against a number of published calibrations from the literature. As most relations are usually calibrated using the Johnson $K$ band we transformed the 2MASS photometry into the Johnson photometric system. In Fig.~\ref{sbcr:comp} we plot our fit and a few calibrations based on interferometric measurements of stellar angular diameters \citep{ker04,cha14,boy14,sal20}. The calibrations based on the interferometry have precision of about 2--3\% for dwarfs in the colour range of our sample. The SBCR derived from eclipsing binary stars and interferometry show very good consistency, with mean offsets below 2\%. Especially, the relation by \cite{ker04} is very consistent with our polynomial fit: in a whole range of $(V\!-\!K)$ colour the differences between the two relations are smaller than 1.5\%. The exception is the relation by \cite{cha14} in a colour range of 0.1--1.0 where there is a systematic offset of about 3\%. The reason of the discrepancy is not clear to us at the moment, but selection effects are a possible explanation. 

We compared also our new SBCR with the linear relation derived for the $(V\!-\!K)$ colour by \citet{gra17}. The \textit{rms} of our new relation is almost three times smaller than the previous one. There is a systematic difference between the previous relations amounting to 1.5\%, which is well within the precisions of the calibrations (2.7\%). The reason for the offset is that in the previous calibration we used less precise and accurate parallaxes from \textit{Hipparcos} and \textit{Gaia} DR1.

\section{Final remarks} \label{fin}
We have presented the results of a detailed analysis of four well-detached eclipsing binary systems. We reached a very high precision in the determination of the fundamental physical parameters of the component stars. This was possible thanks to the high quality of ground- and space-based photometry and spectroscopy we used, and a careful multi-step analysis incorporating multiple consistency checks for the derived parameters.

Some of the analysed systems, for example AL~Ari, may become benchmark stars due the high precision of the derived parameters and the number of different techniques used in their investigation. AL~Dor is a model system for calibration of the SBCR because its components are almost identical. FM~Leo and BN~Scl can be very useful for testing the strength of convective core overshooting in stars for a mass range between 1 and 2 M$_\odot$.

The main result of this paper is a new and very precise calibration of the surface brightness -- colour relations, using {\it Gaia} EDR3 parallaxes. For the first time we achieved it by using only eclipsing binary stars with secure geometric distances. Although in our previous paper \citep{gra17} we used a similar methodology, the relatively low precision of the available parallaxes forced us to use photometric distance priors which led to small systematic biases in the derived angular diameters.

We expect the new SBCRs to be a useful tool for predicting the angular diameters of stars, especially as we calibrated them using all-sky homogenous photometric systems: 2MASS and {\it Gaia}. One of the immediate applications is the determination of the absolute zero-point shift of the {\it Gaia} EDR3 parallaxes in a way similar to the one presented by \cite{gra19} for {\it Gaia} DR2.

In what way could the calibrations be improved? First, the colour range over which the calibrations are applicable could be extended by including in the sample more nearby eclipsing binary systems with early-type components (late B and A type stars) or late-G type components. Second, the uncertainties in the infrared photometry could be driven down via multi-epoch observations with a relative precision of about 1\% in the $J$ and $K$ bands. Third, the sample size could be increased via a re-analysis of a sub-sample of known systems for which new high-quality light curves from K2 or TESS are available. And fourth, a possible parallax bias could be removed by utilising future {\it Gaia} data realeses in which binary motion will be included in astrometric solutions. By achieving all these steps we could expect to reach for the first time a sub-percent precision in the predicted angular diameters of stars.

\begin{acknowledgements}
This work has made use of data from the European Space Agency (ESA) mission
{\it Gaia} (\url{https://www.cosmos.esa.int/gaia}), processed by the {\it Gaia}
Data Processing and Analysis Consortium (DPAC,
\url{https://www.cosmos.esa.int/web/gaia/dpac/consortium}). Funding for the DPAC
has been provided by national institutions, in particular the institutions
participating in the {\it Gaia} Multilateral Agreement.
\\
We are grateful to J.V.~Clausen, B.E.~Helt, and E.H.~Olsen for making their
unpublished $uvby$ photometric data available to us.
\\
The research leading to these results has received funding from the European Research Council (ERC)
under the European Union's Horizon 2020 research and innovation program (grant agreement No 695099)
and from the National Science Center, Poland grants MAESTRO UMO-2017/26/A/ST9/00446 and
BEETHOVEN UMO-2018/31/G/ST9/03050. We acknowledge support from the IdP II 2015 0002 64 and
DIR/WK/2018/09 grants of the Polish Ministry of Science and Higher Education.
\\
W.G.\ also gratefully acknowledges financial support for this work from the BASAL Centro de Astrofisica y Tecnologias Afines
BASAL-CATA (AFB-170002), and from the Millenium Institute of Astrophysics (MAS) of the Iniciativa Cientifica Milenio del Ministerio de Economia,
Fomento y Turismo de Chile, project IC120009. PK and NN acknowledge the support of the French Agence Nationale de la Recherche (ANR),
under grant ANR-15-CE31-0012-01 (project UnlockCepheids).
\\
MT acknowledges financial support from the Polish National Science Center
grant PRELUDIUM 2016/21/N/ST9/03310.
\\
The research leading to these results has (partially) received funding from the KU~Leuven Research Council (grant C16/18/005: PARADISE), from the Research Foundation Flanders (FWO) under grant agreement G0H5416N (ERC Runner Up Project), as well as from the BELgian federal Science Policy Office (BELSPO) through PRODEX grant PLATO.
 \\
This research has made use of the VizieR catalogue access tool, CDS,
 Strasbourg, France (DOI : 10.26093/cds/vizier). The original description
 of the VizieR service was published in 2000, A\&AS 143, 23.
 \\
We used the {\it uncertainties} python package.

\end{acknowledgements}

%\appendix

\label{lastpage}

\end{document}